\input amstex
\documentstyle{amsppt}
\define\Sigmabar{\overline\Sigma}
\define\fcb{\hat\partial}
\define\htop{$\;\widehat{}\;$-topology\;}
\define\hto{$\;\widehat{}\;$-topolog}
\define\hpG{\widehat{\pi_G}}
\define\hXG{\widehat{X_G}}
\define\hX/G{\widehat{X/G}}
\define\inv{^{-1}}
\define\ip{\Cal I\Cal P}
\define\cp{\bigcup\,}
\define\mk{\Bbb L^}
\define\st{\,|\,}
\define\V{\widehat V}
\define\M{\widehat M}
\define\tP{\widetilde P}
\define\tQ{\widetilde Q}
\define\piG{\hat\pi_G}
\define\piGb{\piG^\partial}
\define\<{\langle}
\define\>{\rangle}
\define\ie{{\it i.e.\/}}
\define\R{\Bbb R^}
\define\Sph{\Bbb S^}

\topmatter
\title Discrete Group Actions on Spacetimes: \\
Causality Conditions and the Causal Boundary \endtitle
\rightheadtext {Discrete Group Actions and the
Causal Boundary}
\author
Steven G. Harris
\endauthor
\address
Department of Mathematics, Saint Louis University,
St. Louis, MO 63103, USA\endaddress
\email harrissg\@slu.edu\endemail
\abstract
Suppose a spacetime $M$ is a quotient of a spacetime
$V$ by a discrete group of isometries.  It is
shown how causality conditions in the two
spacetimes are related, and how can one learn about the
future causal boundary on $M$ by studying structures in
$V$.  The relations between the two are particularly
simple (the boundary of the quotient is the quotient of
the boundary) if both $V$ and $M$ have spacelike future
boundaries and if it is known that the quotient of the
future completion of $V$ is past-distinguishing.  (That
last assumption is automatic in the case of $M$ being
multi-warped.) 
\endabstract
\endtopmatter

\head 0: Introduction \endhead

Symmetry conditions on spacetimes are often employed
to simplify various considerations and make important
features stand out.   Such symmetries are often
expressed by the discrete action of a group of
isometries; an example is the fundamental group acting
on the universal cover of a space. 

The impetus for this paper is to isolate and understand
the effect of these kinds of constructions on the causal
structures of spacetimes, particularly in such
properties as strong causality and the causal boundary of
Geroch, Kronheimer, and Penrose \cite{GKP},
\cite{HE} (as topologized in \cite{H2}).  To this effect,
we will look at the situation of a principle (or regular)
covering projection of spacetimes $\pi: V \to M$ given
by a discrete group $G$ of isometries acting freely and
properly discontinuously on $V$: $M = V/G$ and
$\pi(x) = G \cdot x$ is the orbit of $x$ under $G$. 

This context is inspired by the thought of starting
with a spacetime $M$ and letting $V$ be $\widetilde
M$, the universal covering space of $M$, with $G$ the
fundamental group of $M$, $\pi_1(M)$.  But the results
apply just as well for $V = \widetilde M/H$ for
$H$ any normal subgroup of $\pi_1(M)$, with $G =
\pi_1(M)/H$; and some of the results also apply for
a non-principle (non-regular) cover, \ie, with
$V = \widetilde M/H$ for $H$ a non-normal subgroup
of $G$ and $\pi : V \to M$ a covering projection not
given by a group action. 

An important application of the results
of this paper will be an investigation (in a
subsequent paper) of the causal boundary of any
strongly causal static-complete spacetime (static with
the timelike Killing field being complete).  Any such
spacetime has a standard static spacetime as its
universal covering space, and the causal boundary of
standard static spacetimes was given in \cite{H3}. 
Thus, the results of this paper will be useful in
extending the results in \cite{H3} to more general static
spacetimes.

Section 1 shows how causality properties of $M$ and
$V$ are related.  Section 2 explores various
$G$-invariant constructions in $V$ which yield
information about the causal boundary of $M$; much of
that will be discussed in the more general context of
chronological sets.  Section 3 explores the simplified
relations that obtain when $M$ and $V$ are restricted to
have spacelike boundaries.

\head 1: Spacetime Covering Projections and Causality
Conditions
\endhead

Let $V$ be a spacetime, that is to say, a
connected Lorentz manifold with a time-orientation; let
$G$ be a group of time-orientation-preserving
isometries of $V$ acting freely and properly
discontinuously on $V$, that is to say, for every
point $p \in V$ there is a neighborhood $U$ of $p$
such that for every $g \in G$ except the identity
element $e$, $(g \cdot U) \cap U = \emptyset$.  Then $M
= V/G$ is also a connected manifold and inherits a
Lorentz metric and time-orientation from $V$, making
$M$ a spacetime also.  The chronology relation in
either spacetime will be denoted by $\ll$, the
causality relation by $\prec$.  Let $\pi : V \to M$ be
the natural projection, carrying a point $p$ to its
equivalence class, \ie, its orbit under $G$, $G \cdot
p$.  Then $\pi$ is a principle (or regular) covering
projection \ie, the fibre is a group yielding the
deck transformations of the covering projection; and
that is the way in which any principle covering
projection between spacetimes can be constructed.  

As indicated before, we could just have easily begun
with a spacetime $M$ and let $V = \widetilde M/H$,
where $\widetilde M$ is the universal covering space of
$M$ and $H$ is any normal subgroup of $\pi_1(M)$; then
$G = \pi_1(M)/H$ is the fibre-group of the projection.

Somewhat more generally, we could start with a
spacetime $M$ and let $V = \widetilde M/H$ for $H$ any
subgroup of $G = \pi_1(M)$, and let $\pi : V \to M$
be defined by $\pi : H \cdot z \mapsto G \cdot z$ for
$z \in \widetilde M$ (identifying points in $M$ or
$V$ with, respectively, the $G$- or $H$-orbits of
points in $\widetilde M$).  This is a covering
projection (not principle unless $H$ is normal in
$G$), and any covering projection between spacetimes
arises in this manner.  The fibre of the projection is
$G/H$.  We have an equivalence relation $\cong$
defined on $V$ by $p \cong q$ if and only if $p$ and
$q$ are in the same fibre (\ie, map to the same point
by $\pi$); we may call this the fibre equivalence
relation. 

(It is somewhat more awkward to recast this situation
as starting with $V$:  One must start with an
equivalence relation $\cong$ on $V$ with every point
$p$ having a neighborhood $U$ with no two points
of $U$ being related and with $\{r \in V
\;|\; r \cong q \text{ for some } q \in U\}$
consisting of a collection of disjoint open sets each
isometric to $U$.  The more usual approach in
topology is to start with the notion of a covering
projection $\pi : V \to M$.)

The central idea in all these formulations of a
covering projection $\pi : V \to M$ between spacetimes
is that $\pi$ locally be a
time-orientation-preserving isometry. Let us call
such a map a {\it spacetime covering projection\/};
$V$ is the total space, $M$ the base space.  The most
fundamental (and obvious) relation obtaining in such a
context is this:

\proclaim{Proposition 1.1} Let $\pi : V \to M$ be a
spacetime covering projection.  For $p$ and $q$
in $V$, $p \ll q$ implies $\pi(p) \ll \pi(q)$; for $x$
and $y$ in $M$, $x \ll y$ implies for any $p \in
\pi^{-1}(x)$, there is some $q \in \pi^{-1}(y)$ with $p
\ll q$.  Similar results hold for $\prec$.
\endproclaim 

\demo{Proof} If $\bar  c$ is a future-directed timelike
curve in $V$, then $\pi \circ \bar  c$ is a
future-directed timelike curve in $M$.  Thus, a
future-timelike curve from $p$ to $q$ in $V$ projects to
a future-timelike curve from $\pi(p)$ to $\pi(q)$.  

Conversely, a future-directed timelike curve $c$ in $M$
has a unique lift $\bar  c$ to a curve in $V$ starting
from any given point in the pre-image of the
beginning-endpoint of $c$, and $\bar  c$ is a
future-directed timelike curve.  Thus, a
future-timelike curve from $x$ to $y$ in $M$ lifts to a
future-timelike curve from any $p$ sitting above $x$ to
some $q$ sitting above $y$.

Similarly for causal curves and
the causality relation. \qed \enddemo

For a principle covering projection with fibre group
$G$, we can recast this:  For $p$ and $q$ in $V$,
$\pi(p) \ll \pi(q)$ (respectively, $\pi(p) \prec
\pi(q)$) if and only if for some $g \in G$, $p \ll g
\cdot q$ (respectively, $\pi(p) \prec g \cdot \pi(q)$).

The idea of this section is to explore, in respect of
various causality conditions, the relation between
the condition obtaining on the total space and the
base space of a spacetime covering projection.  As a
general rule:  A causality condition on the base space
holds if and only if a somewhat stronger property
holds on the total space.

Surely the first conditions to be examined are those
of chronology and causality.

\proclaim{Proposition 1.2} Let $\pi: V \to M$ be a
spacetime covering projection between spacetimes with
fibre equivalence relation $\cong$. Then
$M$ is chronological (respectively, causal) if and
only if  for all $p, q \in V$, if $p \cong q$, then $p
\not\ll q$ (respectively, $p \not\prec q$).
\endproclaim

\demo{Proof} If $p \cong q$ in $V$, then $\pi(p) =
\pi(q)$; call this point (in $M$) $x$.  Then, by
Proposition 1.1, if $p \ll q$, we have $x \ll x$, so $M$
is not chronological; and if $p \prec q$, we have $x
\prec x$, so $M$ is not causal.

Conversely, suppose for some $x \in M$, $x \ll x$,
\ie, there is a timelike curve $c$ from $x$ to $x$;
say $c: [0,1] \to M$, future-directed, with both
endpoints at
$x$.  Pick a point $p \in \pi^{-1}(x)$.  Then $c$ lifts
to a future-directed timelike curve $\bar  c : [0,1] \to
V$ with $\bar  c(0) = p$ and $\pi(\bar  c(t)) = c(t)$ for
all $t$.  In particular, $q =\bar  c(1)$ lies in
$\pi^{-1}(x)$, so $q \cong  p$; then $p \ll q$. 
Similarly, if $x \prec x$, then $p \prec q$.
\qed \enddemo

It may be worth noting that in the case of a principle
covering projection, with the fibre equivalence
relation given by the action of a fibre group $G$, the
condition on $V$ in Proposition 1.2 amounts to saying
that no point $p$ is chronologically (or causally)
related to any member of its orbit $G \cdot p$.

Strong causality may be the next most widely
applicable causality condition.

\proclaim{Proposition 1.3} Let $\pi: V \to M$ be a
spacetime covering projection. Then $M$ is
strongly causal if and only if for all $p \in V$,
there is a  fundamental neighborhood system $\{U_n\}$
for $p$ such that for each $n$, no causal curve can
have one endpoint in $U_n$ and another endpoint in a
component $U'_n$ of $\pi^{-1}(\pi[U_n])$ unless $U'_n =
U_n$ and the curve remains wholly within $U_n$.

An equivalent condition on the fundamental neighborhood
system $\{U_n\}$ is that for each $n$, no causal curve may
exit and re-enter $U_n$, $\pi$ restricts to a
homeomorphism on $U_n$, and for any two components
$U_n'$ and $U_n''$ of $\pi\inv(\pi[U_n])$, no point of
$U_n'$ is timelike related to any point of $U_n''$.
\endproclaim

\demo{Proof} Suppose $p \in V$ has such a fundamental
neighborhood system $\{U_n\}$; then $\{W_n =
\pi(U_n)\}$ is a fundamental neighborhood system for
$x = \pi(p)$, if we restrict $n$ to be so large that
each $U_n$ is carried homeomorphically by $\pi$ to
its image $W_n$.  Any causal curve
$c$ in $M$ which starts and ends in some $W_n$ lifts
to a timelike curve $\bar  c$ in $V$, beginning in
$U_n$ and ending in some component $U'_n$ of
$\pi^{-1}(\pi(U_n))$.  Then
$U'_n = U_n$ and $\bar  c$ remains within $U_n$; thus,
$c$ does not exit $W_n$.  Therefore, $\{W_n\}$ is the
required neighborhood system showing strong causality
at $x$.

Conversely, suppose strong causality holds at $x \in
M$, \ie,  $x$ has a fundamental neighborhood system
$\{W_n\}$ with no causal curve in $M$ exiting and
re-entering any $W_n$.  Let $p$ be any point in
$\pi^{-1}(x)$.  For $n$ sufficiently large, each
$\pi^{-1}(W_n)$ has exactly one component
$U_n$ containing $p$ with $\pi: U_n \to W_n$ a
homeomorphism; these $\{U_n\}$ form a fundamental
neighborhood of $p$.  For any causal curve
$\bar  c$ in $V$ with one endpoint in $U_n$ and another
endpoint in any component $U'_n$ of $\pi^{-1}(W_n)$,
$c = \pi \circ \bar  c$ is a causal curve in
$M$ with both endpoints in $W_n$.  Then $c$ must
remain wholly within $W_n$; hence, $\bar  c$ remains
wholly within $U_n$, whence $U'_n = U_n$. 

The condition in the second paragraph is equivalent,
since a causal curve exiting one component and entering
another implies a timelike relation between points of the
components. \qed \enddemo

In the case of a principle covering projection with
fibre group $G$, the condition on $V$ above says that
each point has a fundamental neighborhood system
$\{U_n\}$ with no causal curve exiting and re-entering
any $U_n$ and no points of $U_n$ and $g\cdot U_n$ being
timelike related unless $g = e$.

The next result concerns global hyperbolicity.  We
will use $J(x,y)$ to denote $J^+(x) \cap J^-(y)$,
where $J^+$ and $J^-$ denote causal future and past,
respectively (recall that global hyperbolicity is
equivalent to strong causality plus each
$J(x,y)$ being compact).  For this result, very
different arguments will be used in the two
directions:  Arguing from the total space to the base
space, the sort of direct methods employed above work
for consideration of the spaces $J(x,y)$; but for
arguing from the base space to the total space, it is
far easier to use the equivalent condition (\cite{HE},
section 6.6) that the spacetime has a Cauchy surface.

\proclaim{Proposition 1.4} Let $\pi : V \to M$ be a
spacetime covering projection. Then $M$ is globally
hyperbolic if and only if 
\roster
\item $V$ is globally hyperbolic,
\item every point $p \in V$ has a fundamental
neighborhood system as in Proposition 1.3, and 
\item for any $p \in V$, for all $q \gg p$, $J^+(p)
\cap \pi^{-1}(\pi(q))$ is finite.
\endroster
\endproclaim

\demo{Proof} Suppose that $M$ is globally hyperbolic. 
That means it has a Cauchy surface $\Sigma$.  Using a
technique from \cite{GH} (Lemma 4.1), we will see that
$\Sigma$ gives rise to a Cauchy surface $\Sigmabar$
in $V$, showing that $V$ also is globally hyperbolic
(but we must fill in the gap in \cite{GH}, which
failed to show why $\Sigmabar$ must be connected).

Let $\Sigmabar = \pi^{-1}(\Sigma)$; then $\Sigmabar$
is a topological spacelike hypersurface in $V$ (just
as $\Sigma$ is in $M$), and every endless causal curve
$\bar  c : (-\infty, \infty) \to V$ intersects
$\Sigmabar$ exactly once:  The projection $c = \pi
\circ \bar  c : (-\infty, \infty) \to M$ is an endless
causal curve so it intersects
$\Sigma$ exactly once.  Since for any $t$, $c(t) \in
\Sigma$ if and only if $\bar  c(t) \in \Sigmabar$, this
means there is exactly one $t_0$ with $\bar  c(t_0) \in
\Sigmabar$.  Thus, all we need to conclude that
$\Sigmabar$ is a Cauchy surface is that it be
connected.

Consider any point $x \in \Sigma$ and two points $p$
and $q$ in $\pi^{-1}(x)$. Let $\bar \sigma$ be a curve
in $V$ from $p$ to $q$ (we are assuming that $V$ is
connected); then $\sigma = \pi \circ \bar \sigma$ is a
closed curve in $M$ containing $x$.  There is a
homotopy $h$, fixing $x$, from $\sigma$ to a closed
curve $\sigma'$ lying wholly in $\Sigma$ (this is
because, topologically, $M = \Bbb R\times \Sigma$---see
\cite{HE}, Proposition 6.6.8 ; if, say, $\sigma(t) =
(s(t), y(t))$ with $x = (0,y(0))$, then let $h(t,u) =
(us(t), y(t))$, $0 \le u \le 1$).  Then, by the
homotopy lifting property of covering projections,
there is a lift of $h$ to a homotopy $\bar h$ in $V$,
and $\bar h$ is a homotopy from $\bar \sigma$ to some
curve $\bar \sigma'$, with the same endpoints as
$\sigma$, lying wholly in $\Sigmabar$.  Therefore, $p$
and $q$ are in the same component of $\Sigmabar$.

This shows that all the elements of $\Sigmabar$
sitting over the same point in $\Sigma$ are in the
same component of $\Sigmabar$.  Let $C$ be any
such component; then $\pi(C)$ is open in $\Sigma$
(since $\pi : \Sigmabar \to \Sigma$ is a local
homeomorphism) and connected, hence, a component of
$\Sigma$.  Since $\Sigma$ is connected, $\pi(C)$ must
be all of $\Sigma$, so $C$ includes points sitting
above each point of $\Sigma$.  That means $C$ is all
of $\Sigma$, \ie, $\Sigmabar$ is connected.  This
shows $V$ is globally hyperbolic.

Since $M$ is strongly causal (being globally
hyperbolic), we know from Proposition 1.3 that $V$ must
satisfy condition (2).

For condition (3), suppose there is a point $p \in V$
above a point $x \in M$ and that there is an infinite
number of points $\{q_n\}$ in $V$, all above the same
point $y \in M$, with all $q_n \succ p$.  Then we have
future-causal curves $\bar  c_n$ from $p$ to $q_n$;
each yields a future-causal curve $c_n = \pi \circ
\bar  c_n$ from $x$ to $y$.  If any two of these curves
in $M$, $c_n$ and $c_m$, were homotopic, then their
lifts $\bar  c_n$ and $\bar  c_m$ in $V$ would have the
same endpoints; but they don't, since $q_n$ and $q_m$
are distinct for $n \neq m$.  Thus, the curves
$\{c_n\}$ represent distinct homotopy classes in $M$.
 
However, $M$ being globally hyperbolic, the causal
curves from $x$ to $y$ form a compact set in the
compact-open topology (\cite{HE}, Proposition 6.6.2);
thus, there is a future-causal curve $c$ from $x$ to
$y$ and a subsequence $\{c_{n_k}\}$ approaching $c$ in
this topology.  In particular, for any neighborhood
$U$ in $M$ of the image of $c$, an infinite number of
the $\{c_n\}$ lie in $U$.  Let us parametrize $c$ on
$[0,1]$; then we can choose $U$ to be a small tubular
neighborhood of the form $(-\epsilon,1+\epsilon)
\times B^{n-1}$ (where $n$ is the dimension of $M$ and
$B$ denotes the Euclidean ball about the origin
$\bold 0$), with $x$ corresponding to $(0,\bold 0)$
and $y$ to $(1,\bold 0)$.  Then any causal curve $c'$
in $U$ from $x$ to $y$ must be homotopic to $c$ (with
$U$ sufficiently small, $c'$ must lie in $[0,1]
\times B^{n-1}$, hence, be parametrizable as
$(t,z(t))$ for $z(t) \in B^{n-1}$).  Thus, for any $n$
and $m$ with $c_n$ and $c_m$ both lying in $U$, both
are homotopic to $c$, hence, to one another---a
contradiction.  Therefore, condition (3) is true.

Conversely, suppose conditions (1), (2), and (3) are
true.  By Proposition 1.3, we know that $M$ is strongly
causal; we just need to show that for every  $x$ and
$y$ in $M$, $J(x,y)$ is compact.  Let $p$ be any
lift of $x$ to $V$.  For any point $z \in J(x,y)$,
there is a future-causal curve $c$ from $x$ to $y$
passing through $z$; we can lift that to a
future-causal curve $\bar  c$ in $V$ starting at $p$
and including a point $r$ sitting above $z$.  Then
$\bar  c$ terminates at some point $q$ sitting above
$y$, and we have $r \in J(p,q)$.  Therefore, every
point in $J(x,y)$ can be expressed as the projection
of a point in $J(p,q)$ for some $q \in \pi^{-1}(y)$. 
But there are only finitely many points $\{q_1, ...,
q_k\}$ sitting above $y$ for which each $J(p,q_i)$ is
non-empty.  Thus, $J(x,y) = \pi\left(\bigcup_{i \le k}
J(p,q_i) \right)$; since each $J(p,q_i)$ is compact
($V$ being globally hyperbolic) and the continuous
image of a compact set is compact, this means $J(x,y)$
is compact. \qed \enddemo

Clearly, condition (3) can be replaced by a similar
statement involving $J^-$.  In the case of a principle
covering projection with fibre group $G$, condition (3)
says that for any $p$ and $q$ in $V$,  $J^+(p) \cap ( G
\cdot q)$ is finite.

It's not entirely clear that condition (3) is needed
in Proposition 1.4, \ie, although it is implied by $M$
being globally hyperbolic, it may be that it is a
consequence of conditions (1) and (2) anyway.  Since
we can always formulate the context in terms of group
actions on the universal covering space, the question
comes down to this:  If $V$ is a globally hyperbolic
spacetime acted on freely and properly discontinuously
by a group $G$ of time-orientation-preserving
isometries, is it possible that there is a point $p$
and a point $q$ with $G \cdot q$ having an
infinite number of elements in $J^+(p)$?

\head
2: Discrete Group Actions and the Future Causal Boundary
\endhead

The discussion in this section will focus on the chief
aspects of the causal boundary of
\cite{GKP}---specifically, the future causal
(pre-)boundary as formulated first in \cite{H1} (for
categorical construction) and then in \cite{H2} (for
topology).  As shown in \cite{H1}, the full causal
boundary has innate difficulties that make it resistant
to categorical formulation.  But by restricting to the
future causal boundary we can formulate future-completion
in a manner that applies to spacetimes as merely one
example of a much broader category of  ``chronological
sets"; then the process of future-completion (adding the
future causal boundary) within this larger category is
fully categorical (\ie, functorial) and also universal
(hence, unique) in the sense of category theory.  

The topology of the causal boundary as formulated in
\cite{GKP} cannot be implemented with just the future
causal boundary (and it has severe problems of its own). 
But a topology for any chronological set was developed in
\cite{H2}, and this topology was shown also to have
categorical and universal properties, at least in the
case of spacelike boundaries.

The intent of this section is to explore the
future causal boundary for spacetimes with discrete group
actions.  However, the constructions naturally lend
themselves to exposition within the broader category of
chronological sets.  As it turns out, this is a good
thing, for the slickest presentation of the results lies
in a construction with chronological sets that takes us
far afield from spacetimes, even when that is what we
start with.  A return to constructions more narrowly
focused in spacetimes will be the subject of section 3.
Subsection 2.1 will look purely at basic properties of
chronological sets, while Subsection 2.2 will consider
aspects of topology derived from the chronology
relation.

\subhead 2.1: Discussion of the Chronological Relation
\endsubhead  \medskip

We will make use of some definitions and results from
\cite{H1}, recalled here:

A {\it chronological set\/} is a set
$X$ together with a relation $\ll$ (the {\it chronology
relation\/}) obeying the following:
\roster
\item $\ll$ is transitive: $x \ll y$ and $y \ll z$
implies $x \ll z$
\item $\ll$ is non-reflexive:  for all $x$, $x \not\ll
x$
\item There are no isolates:  for all $x$, the past of
$x$, $I^-(x) = \{y \,|\, y \ll x\}$, or the future of
$x$, $I^+(x) = \{y \,|\, x \ll y\}$, (or both) is
non-empty.
\item $X$ is separable:  There is a countable subset $S_0
\subset X$ such that for $x \ll y$, there is some $s
\in S_0$ with $x \ll s \ll y$.  (We can say $S_0$ is
``chronologically dense" in $X$.)
\endroster 
A {\it past set\/} in $X$ is a subset $P \subset X$
such that $P = I^-[P]$ (where $I^-[A] = \bigcup_{a \in
A} I^-(a)$); any set of the form $I^-[A]$ is a past
set.  A past set is {\it
indecomposable\/} if it is not the union of two past
sets which are proper subsets; a past set $P$ is
indecomposable if and only if for any $x$ and $y$ in
$P$, there is some $z \in P$ with $x \ll z$ and $y \ll
z$.   A {\it future chain\/} is
a sequence $\{x_n \,|\, n \ge 1\}$ with $x_n \ll
x_{n+1}$, all $n$ (this plays the role of a timelike
curve in a spacetime).  For any future chain $c$,
$I^-[c]$ is an IP (indecomposable past set), and for
every IP $P$, there is a future chain $c$ in $P$ such
that $P = I^-[c]$.  A point $x \in X$ is a
{\it future limit\/} of the future chain $c$ if
$I^-(x) = I^-[c]$ (in a spacetime, this is the same as
the topological limit of the sequence); a future limit
of a future chain is unique if $X$ is {\it
past-distinguishing\/} ($I^-(x) = I^-(y)$ implies $x =
y$). $X$ is {\it future-complete\/} if every future
chain has a future limit.  A map
$f : X \to Y$ between chronological sets is {\it
future-continuous\/} if $x_1 \ll x_2$ implies $f(x_1)
\ll f(x_2)$ and if $x$ being a future limit of the future
chain $\{x_n\}$ implies $f(x)$ is a future limit of
$\{f(x_n)\}$.  Let $\fcb(X) = \{P \subset X
\,|\, P \text{ is an IP and } P \text{ is not }
I^-(x) \text{ for any } x\in X\}$ (the {\it future
chronological boundary\/} of $X$) and let $\hat X = X
\cup \fcb(X)$ (the {\it future completion\/} of
$X$), and extend $\ll$ to $\hat X$ via (for $x \in X$,
$P \in \fcb(X)$, $Q \in \fcb(X))$
\roster
\item $x \ll P$ iff $x \in P$ 
\item $P \ll x$ iff $P \subset I^-(w)$ for some $w \ll
x$
\item $P \ll Q$ iff $P \subset I^-(w)$ for some $w \in
Q$.
\endroster
Then $\hat X$ is a future-complete chronological set
(past-distinguishing if $X$ is), and the inclusion
$\iota_X: X \to \hat X$ is future-continuous. 

$X$ is {\it past-determined\/} if $x \ll y$ whenever
$I^-(x)$ is non-empty and $I^-(x)
\subset I^-(w)$ for some $w \ll y$ (example: globally
hyperbolic spacetimes are past-determined);
future-completion preserves being past-determined.  For
any future-continuous map $f : X \to Y$ with
$Y$ past-determined and past-distinguishing, there is a
unique future-continuous map $\hat f: \hat X \to \hat
Y$ extending $f$ (\ie, $ \hat f \circ \iota_X = \iota_Y
\circ f$).  This amounts to a functor
$\;\widehat{}\;$ on the category of
past-determined, past-distinguishing chronological
sets (morphisms being the future-continuous maps), with
target the sub-category of future-complete sets; it
obeys the universality property that for any $f: X \to
Y$ in the category with $Y$ future-complete,
there is a unique extension of $f$ to $\hat X$.  For
chronological sets which are not past-determined, we
can apply the {\it past-determination\/} functor, which
adds some $\ll$ relations:  $X^p$ is the set $X$ with $x
\ll y$ defined to hold if and only if, within $X$, $x
\ll y$ or, with $I^-(x)$ non-empty, $I^-(x) \subset
I^-(w)$ for some $w \ll y$.  Then $X^p$ is a
past-determined chronological set (past-distinguishing
and future-complete if $X$ is).  The GKP future causal
boundary of a strongly causal spacetime $X$, added to
$X$, results in precisely $(\hat X)^p$, and this is
canonically isomorphic to $\widehat{X^p}$.  (That
canonical isomorphism---a natural equivalence in the
categorical sense---is the only really difficult part of
the whole enterprise; see \cite{H1}, Proposition 13.)

\medskip

What we are interested in here is a chronological set
$X$ with an action from a group $G$.  We need the
action to consist of isomorphisms of the chronology
relation, \ie, $g \cdot x \ll g \cdot y$ iff $x \ll
y$ (for the covering maps of the previous section, we
also needed the action to be free---for any $g \in G$ with
$g \neq e$, for all $x \in X$, $g \cdot x \neq x$---but
that is not required for the matters in this section).  We
need to know, first of all, when it is that
$X/G$, the set of $G$-orbits, inherits the structure of a
chronological set; for $G \cdot x$ representing the
$G$-orbit of a point $x$, we want $G \cdot x \ll G \cdot
y$ if and only if $x \ll g \cdot y$ for some $g \in
G$.  (We will sometimes need to think of $G
\cdot x$ as a single element of $X/G$ and sometimes need
to think of it as the several elements in the orbit of
$x$; the context should make it clear.)  Then we want to
find a way to describe $\widehat{X/G}$ and
$\fcb(X/G)$ in terms of constructions on $X$.

The first matter is easy:

\proclaim{Proposition 2.1} Let $X$ be a chronological
set with a group action via chronological
isomorphisms from a group $G$.  Then the set $X/G$
inherits the structure of a chronological set if and
only if for all $g \in G$ and $x \in X$, $x \not\ll g
\cdot x$.  When this obtains, the projection
$\pi : X \to X/G$ is future-continuous.
\endproclaim

\demo{Proof}  $X/G$ is necessarily transitive ($x
\ll g \cdot y$ and $y \ll h \cdot z$ yield $g \cdot y
\ll g \cdot (h \cdot y) = (gh) \cdot z$, so $x \ll (gh)
\cdot z$).  Since $x \ll y$ implies $G \cdot x \ll G
\cdot y$, if $x$ has a non-empty past (or future), so
does $G \cdot x$.   If $S_0$ is the countable set making
$X$ separable, then $\pi[S_0] = \{G \cdot s \,|\, s \in
S_0\}$ is a countable collection of elements of $X/G$ and
serves the same purpose in that set.  The only thing that
could go wrong is non-reflexivity:  $G \cdot x \ll G
\cdot x$ occurs if and only if $x \ll g \cdot x$ occurs. 

Assume now that $x \ll g \cdot x$ never occurs.  We
already know that $x \ll y$ implies $\pi(x) \ll
\pi(y)$.  Suppose $c = \{x_n\}$ is a future chain in $X$
and $I^-(x) = I^-[c]$; we want to show that
$I^-(\pi(x)) = I^-[\pi(c)]$.  We have $\pi(y) \ll
\pi(x)$ means $y \ll g \cdot x$ for some $g$, whence
$g^{-1} \cdot y \ll x$, so $g^{-1} \cdot y \ll x_n$ for
some $n$, so $y \ll g \cdot x_n$, \ie, $\pi(y) \in
I^-[\pi(c)]$.  Conversely, $\pi(y) \in I^-[\pi(c)]$
means $y \ll g \cdot x_n$ for some $n$ and some $g$, so
$g^{-1} \cdot y \ll x_n$, so $g^{-1} \cdot y \ll x$, so
$y \ll g \cdot x$ and $\pi(y) \ll \pi(x)$.
\qed \enddemo

For $X$ a chronological set, we will say that a group
$G$ has a {\it chronological action\/} on $X$ if
it acts on $X$ through chronological
isomorphisms and in accord with the property in
Proposition 2.1.  Then we can express the chronology
relation on $X/G$ in terms of $X$ and the projection
$\pi: X \to X/G$ thus:  For $p$ and $q$ in $X/G$, $p
\ll q$ if and only if some (hence, every) element of
$\pi^{-1}(p)$ is in the past of some element of
$\pi^{-1}(q)$, which happens if and only if some (hence,
every) element of $\pi^{-1}(q)$ lies in the future of some
element of $\pi^{-1}(p)$.  For a subset $A \subset X$, we
then have $\pi[I^-[A]] = I^-[\pi[A]]$. (The convention
with brackets vs. parentheses employed here is that for a
function $f: X \to Y$, brackets are used to denote the
application of $f$ to a subset of $X$, \ie, for $A
\subset X$, $f[A] = \{f(x) \,|\, x \in A\}$.)  The
reason is thus:  If $x \ll a \in A$, then $\pi(x)
\ll \pi(a)$; and if $p \ll \pi(a)$ for $a \in A$, then for
some $x \in \pi^{-1}(p)$, $x \ll a$.  Therefore, we can
characterize the chronological set $X/G$ by
$I^-_{X/G}(\pi(x)) = \pi[I^-_X(x)]$, since
$I^-_{X/G}(\pi(x)) = \pi[I^-_X[G \cdot x]] = \pi[\cp G
\cdot I^-_X(x)]$ (for $S \subset X$, $G \cdot S$ = $\{g
\cdot S \,|\, g \in G\}$, will denote the set of
$G$-orbits of $S$, so $\cp G \cdot S$ denotes the set of
points of $X$ in all those $G$-orbits).

Let $X$ be a chronological set with a
chronological action from a group $G$.  To investigate
the IPs and the future chronological boundary in $X/G$,
we need to look at something a bit different in $X$: 
We will call a subset $A \subset X$ $G$-{\it
invariant} if $\cp G \cdot A = A$. 
Note that $\cp G \cdot I^-[A] = I^-[\cp G \cdot A]$ (for
$a \in A$ and $x \ll a$, we have $g \cdot x \ll g \cdot
a$, so $g \cdot x \in I^-[\cp G \cdot A]$; for $x \ll
g \cdot a$, we have $x = g \cdot (g^{-1} \cdot x)$ and
$g^{-1} \cdot x \ll a$, so $x \in \cp G \cdot I^-[A]$);
hence, a $G$-invariant past set is anything of either of
these forms.  A $G$-{\it indecomposable past set\/} (or a
GIP, for ``group-indecomposable past set", even if another
letter is used to name the group) will be a
$G$-invariant past set which is not the union of two
$G$-invariant past sets which are proper subsets. 
Finally, let the {\it future chronological
$G$-boundary\/} of
$X$, denoted $\fcb_G(X)$, consist of
$\{\text{GIPs } A
\,|\, \text{ for all } x \in X, A \neq G \cdot I^-(x)\}$.

GIPs will be the basic building blocks for our
analysis of $G$-invariant structures in $X$ as they
reflect the items of interest in $X/G$, as
established by the following lemma.  (It will develop
that $\fcb_G(X)$ models $\fcb(X/G)$.)  

\proclaim{Lemma 2.2} Let $X$ be a chronological
set with a chronological action from a group
$G$ with $\pi: X \to X/G$ the natural projection.  For
every GIP $A$ in $X$, $\pi[A]$ is an IP in $X/G$, and for
every IP $P$ in $X/G$, $\pi^{-1}[P]$ is a GIP in $X$. 

Furthermore, an IP $P$ in $X/G$ is an element of
the future chronological boundary $\fcb(X/G)$ if and
only if $\pi^{-1}[P]$ is an element of the future
chronological $G$-boundary $\fcb_G(X)$.
\endproclaim

\demo{Proof}  Since for $x \in X$, $I^-(\pi(x)) =
\pi[I^-(x)]$, we have for any $A \subset X$, $I^-[\pi[A]]
= \pi[I^-[A]]$.  Similarly, for any $p \in X/G$, we have
$I^-[\pi^{-1}(p)] = \pi^{-1}[I^-(p)]$, so for any $P
\subset X/G$, we have $I^-[\pi^{-1}[P]] =
\pi^{-1}[I^-[P]]$. 

Now let $A$ be a past set in $X$; then $I^-[\pi[A]] =
\pi[I^-[A]] = \pi[A]$, showing $\pi[A]$ is a past set. 
And if $P$ is a past set in $X/G$, then $I^-[\pi^{-1}[P]]
= \pi^{-1}[I^-[P]] = \pi^{-1}[P]$, showing $\pi^{-1}[P]$
is a past set; furthermore, as is evident, $\pi^{-1}[P]$
is $G$-invariant.

Now let $A$ be a past set in $X$ with $\pi[A]$
decomposable into proper past sets, $\pi[A] = P_1 \cup
P_2$; then $A = \pi^{-1}[P_1] \cup
\pi^{-1}[P_2]$, and each $\pi^{-1}[P_i]$ is a
$G$-invariant past subset, as well as a proper subset of
$A$.  In particular, if $A$ is a GIP in $X$, then $\pi[A]$
is an IP in $X$.

Let $P$ be a past set in $X/G$ with $\pi^{-1}[P]$
decomposable into proper $G$-invariant past sets
$\pi^{-1}[P] = A_1 \cup A_2$; then $P = \pi[A_1] \cup
\pi[A_2]$, and each $\pi[A_1]$ is a past set, as well as
a proper subset of $P$.  Thus, if $P$ is an IP in $X/G$,
then $\pi^{-1}[P]$ is a GIP in $X$. 

An IP $P$ in $X/G$ is an element of $\fcb(X/G)$ if
and only if for no $p \in X/G$ is $P = I^-(p)$.  We have
$P = I^-(p)$ implies $\pi^{-1}[P] = \pi^{-1}[I^-(p)] =
I^-[\pi^{-1}(p)]$; picking any $x \in \pi^{-1}(p)$, this
yields $\pi^{-1}[P] = I^-[G \cdot x] = G \cdot I^-(x)$. 
On the other hand, if $\pi^{-1}[P] = G \cdot I^-(x)$,
then $P = \pi[G \cdot I^-(x)] = \pi[I^-(x)] =
I^-(\pi(x))$. \qed \enddemo

Thus we know that GIPs are the items of interest in $X$
for analyzing $X/G$.  But what are these new creatures,
and how do they work?  It turns out they are not new at
all, but just the IPs of a new chronological set, $X_G$,
with $\fcb_G(X) = \fcb(X_G)$.

For any chronological set $X$ with a free chronological
action from a group $G$, let us define a new relation
$\ll_G$ on $X$, the $G$-{\it expansion\/} of the
chronological relation $\ll$ on $X$, via $x \ll_G y$ if
and only if for some $g \in G$, $x \ll g \cdot y$, \ie,
if and only if $\pi(x) \ll \pi(y)$ in $X/G$.  Let $X_G$
denote the partially ordered set $(X, \ll_G)$, called the
$G$-{\it expansion\/} of $(X, \ll)$ (when no confusion
results, we will let $X$ stand for $(X, \ll)$, as
before).  

\proclaim{Proposition 2.3} Let $X$ be a chronological
set with a chronological action from a group
$G$. 
\roster
\item"(a)" The $G$-expansion $X_G$ is a chronological set.
\item"(b)"  The natural injection $i_G: X \to X_G$  and
natural projection $\pi_G : X_G \to X/G$ (respectively
the same as the identity map and $\pi$ on the set-level)
are future-continuous.
\item"(c)" The GIPs of $X$ are precisely the IPs of $X_G$.
\item"(d)" The GIPs of $X$ which are elements of
$\fcb_G(X)$ are precisely the IPs of $X_G$ which are
elements of $\hat\partial(X_G)$, \ie, $\fcb_G(X) =
\fcb(X_G)$.
\endroster 
\endproclaim

\demo{Proof} All the properties of a chronological set
follow directly from Proposition 2.1, as does the
future-continuity of $\pi_G$. 
 
Let $I^-_G$ denote the past in $X_G$.  We know that for $x
\in X$, $I^-_G(x) = G \cdot I^-(x)$; then for $A \subset
X$, $I^-_G[A] = G \cdot I^-[A]$. 

For the future-continuity of $i_G$, consider $x$ a future
limit of a future chain $c$ in $X$, \ie, $I^-(x) =
I^-[c]$.  Then $I^-_G(x) = G \cdot I^-(x) = G \cdot
I^-[c] = I^-_G[c]$. 

The $G$-invariant past sets in $X$ are precisely the sets
of the form $G \cdot I^-[A]$; since $G \cdot I^-[A] =
I^-_G[A]$, these are precisely the past sets of $X_G$. 
Thus, indecomposability in the one form amounts to
indecomposability in the other form, yielding the GIPs
of $X$ the same as the IPs of $X_G$.  A GIP which has the
form $G \cdot I^-(x)$ is precisely one which has the form
$I^-_G(x)$. \qed \enddemo 

The $G$-expansion $X_G$ must be treated with care, as it
is a rather anomalous chronological set even if $X$ is
very well-behaved (such as being a strongly causal
spacetime):  Past-distinguishing fails widely in $X_G$,
as for every $x \in X$, for all $g \in G$, $I_G^-(g \cdot
x) = I_G^-(x)$.

Just as IPs in spacetimes can be characterized as the
pasts of timelike curves, and IPs in chronological sets
can be characterized as the pasts of future chains,
GIPs in chronological sets with a chronological
$G$-action can be characterized as the pasts of what might
be thought of as ``$G$-invariant future chains" (in
spacetimes: ``$G$-invariant timelike curves", in analogous
formation).  What this really means is the $G$-orbit of a
future chain; this will be written as $G \cdot c$ for
convenience, though it may be argued that $\cp G \cdot c$
is more proper.

\proclaim{Proposition 2.4} Let $X$ be a chronological
set with a chronological action from a group
$G$.  For any future chain $c$ in $X$, $I^-[G \cdot c]$ is
a GIP, and every GIP can be written in that form.
\endproclaim

\demo{Proof}  Let $c$ be a future chain in $X$; then
$I^-[G \cdot c] = \cp G \cdot I^-[c] = I^-_G[c]$.  Since
$c$ is also a future chain in $X_G$, $I^-_G[c]$ is an IP
in
$X_G$, hence, a GIP in $X$.  

Any GIP $A$, being an IP in $X_G$, can be written as
$I_G^-[c]$ for some future chain $c =\{x_n\}$ in $X_G$. 
Let $x'_1 = x_1$.  Suppose we've defined $x'_1 \ll
\cdots \ll x'_n$ with $x'_i = g_i \cdot x_i$ for some
$g_i \in G$.  We have $x_n \ll_G x_{n+1}$, so $x_n \ll
g \cdot x_{n+1}$ for some $g \in G$; then $x'_n = g_n
\cdot x_n \ll g_ng\cdot x_{n+1}$, and let $x'_{n+1} =
g_ng \cdot x_{n+1}$.  This defines $c' =\{x_n\}$,
a future chain in $X$, with $G \cdot c' = G \cdot c$. 
Then $A = I^-_G[c] = \cp G \cdot I^-[c] = I^-[G \cdot c] =
I^-[G \cdot c']$.   
\qed \enddemo

\proclaim{Corollary 2.5} For any IP $P$ in $X$, $\cp G
\cdot P$ is a GIP, and every GIP can be so expressed.
\endproclaim

\demo{Proof} Any IP can be written as $I^-[c]$ for
some future chain $c$; then $\cp G \cdot I^-[c] = I^-[G
\cdot c]$, which is a GIP by Proposition 2.4. 
Conversely, by Proposition 2.4, any GIP can written as
$I^-[G \cdot c]$ for some future chain
$c$, and $I^-[G \cdot c] = \cp G \cdot I^-[c]$;
$I^-[c]$ is an IP. \qed \enddemo

Following on from Proposition 2.3(b), we can use $\pi_G$
to relate $\fcb_G(X)$, viewed as
$\fcb(X_G)$, to $\fcb(X/G)$ in a direct fashion: 
From $\pi_G : X_G \to X/G$ we can construct the
extension $\hpG : \hXG \to \hX/G$ (recall that $\hat Y$
denotes $Y \cup \fcb(Y)$, the future completion of $Y$). 
It was stated in the introductory portion of this section
that a future-continuous map $f: Z \to Y$ yields a
future-continuous future-completion $\hat f: \hat Z \to
\hat Y$ when $Y$ is past-determined and
past-distinguishing---and $X/G$ might not be
past-distinguishing, even if $X$ is.  But Proposition 6 of
\cite{H1} actually allows for looser conditions.

\proclaim{Proposition 2.6} Let $X$ be a chronological
set with a chronological action from a group
$G$.  The natural projection from the $G$-expansion
$\pi_G : X_G \to X/G$ has a future-continuous extension
to the future-completions, $\hpG : \hXG \to \hX/G$.  It
has these further properties:
\roster
\item For any $\alpha$ and $\beta$ in $\hXG$,
$\hpG(\alpha) \ll \hpG(\beta)$ iff $\alpha \ll_G \beta$.
\item $\hpG$ restricts on the boundaries to an isomorphism
of chronology relations and of the subset
relation, $\bar\pi_G : \fcb(X_G) \cong \fcb(X/G)$.  
\endroster
\endproclaim

\demo{Proof} Proposition 6 of \cite{H1} states that in
order for a future-continuous map $f: Z \to Y$ to have a
future-continuous extension $\hat f : \hat Z \to \hat
Y$, one doesn't really need that $Y$ be
past-distinguishing, but only that for any future chain
$c$ in $Z$ which generates an element of $\fcb(Z)$ (\ie,
$c$ has no future limit in $Z$), there not be two
distinct future limits of $f[c]$ in $Y$.  The map $\hat
f$ is then defined thus:  For $z \in Z$, $\hat f(z) =
f(z)$; for $c$ generating an element $P$ of $\fcb(Z)$, if
$f[c]$ has a (single) future limit $y \in Y$, then $\hat
f(P) = y$; and for $c$ generating $P \in \fcb(Z)$ with
$f[c]$ having no future limit in $Y$, $\hat f(P) =
I^-[f[c]]$, an element of $\fcb(Y)$.

So consider a future chain $c$ in $X_G$.  Suppose
$\pi_G[c]$ has a future limit $p = \pi(x)$ in $X/G$;
then $I^-(\pi(x)) = I^-[\pi[c]]$.  But $I^-(\pi(x)) =
\pi[I^-(x)]$ and $I^-[\pi[c]] = \pi[I^-[c]]$ (as
mentioned in the proof of Lemma 2.2), so we have
$\pi[I^-(x)] = \pi[I^-[c]]$; this implies $\cp G \cdot
I^-(x) = \cp G \cdot I^-[c]$.  But the first of these is
$I^-_G(x)$, the second $I^-_G[c]$; thus, $x$ is a future
limit of $c$ in $X_G$.  Therefore, no future chain
generating a future boundary element of $X_G$ can have
even one future limit for its image in $X/G$, much less
two.

This allows us to define a map $\hpG : \hXG \to \hX/G$;
but Proposition 6 of \cite{H1} uses past-determination of
the codomain to show this map preserves the chronology
relation, and we aren't insisting that $X/G$ be
past-determined.  So we must consider this issue
separately:

For $f : Z \to Y$, past-determination of $Y$ is used only
in showing that for $P \in \fcb(Z)$ and $z \in Z$ with $P
\ll z$ and $\hat f(P) \in Y$, then $\hat f(P) \ll f(z)$
(if $\hat f(P) \in \fcb(Y)$, then $\hat f(P) \ll f(z)$
follows anyway, irrespective of past-determination).  So
consider $P \in \fcb(X_G)$; the import of what was just
shown was that $\hpG(P)$ lies in $\fcb(X/G)$ and not in
$X/G$ (\ie, if the future chain $c$ generates $P$ in
$X_G$, then $\pi[c]$ has no future limit in $X/G$).

Preserving the chronology relation was the only issue;
past-determination and past-distinguishing play no
further role in showing future-continuity. 

For property (1), consider $x$ and $y$ in $X$, $A = \cp G
\cdot P$ and $B = \cp G \cdot Q$ in $\fcb(X_G)$, where $P$
and $Q$ are in $\fcb(X)$, generated by future chains $c$
and $d$ in $X$; we know that $\hpG(A)$ is an element of
$\fcb(X/G)$, specifically $I^-[\pi[c]]$.   We consider
$\alpha \ll_G \beta$ for $\alpha$ and $\beta$
successively being $x$ and $y$, $x$ and $A$, $A$ and $y$,
and $A$ and $B$:

We have $x \ll_G y \iff \pi(x) \ll \pi(y)$.  We have $x
\ll_G A \iff x \in A \iff \pi(x) \in \pi[A] = \pi[P] =
\pi[I^-[c]] = I^-[\pi[c]]$, \ie, $\pi(x) \ll \hpG(A)$. 
We have $A \ll_G y  \iff A \subset I^-_G(w)$ for some $w
\ll_G y \iff \cp G \cdot I^-[c] \subset \cp G \cdot
I^-(w)$ for some $w \ll_G y \iff \pi[I^-[c]] \subset
\pi[I^-(w)]$ for some $w$ with $\pi(w) \ll \pi(y) \iff
\hpG(A) = I^-[\pi[c]] = \pi[I^-[c]] \subset \pi[I^-(w)] =
I^-(\pi(w))$ for some $\pi(w) \ll \pi(y) \iff \hpG(A) \ll
\pi(y)$.   We have $A \ll_G B \iff A \ll_G x \ll_G B$ for
some $x \in X$, which, by the previous results, is
equivalent to $\hpG(A) \ll \hpG(B)$.

For property (2), in light of property (1), we need only
show that $\hpG$ is a bijection from $\fcb(X_G)$ to
$\fcb(X/G)$ that preserves subsets in both directions.  We
already know that $\hpG$ takes boundary elements to
boundary elements.  To show it is onto the boundary,
consider any $P \in \fcb(X/G)$ generated by a future chain
$\sigma = \{p_n\}$.  For each $n$, pick any
$x_n \in \pi^{-1}(p_n)$; then $c = \{x_n\}$ is a future
chain in $X_G$ (property (1)).  If $c$ has a future limit
$x$ in $X_G$, then (again by property (1)) $\pi(x)$ is a
future limit of $\sigma$ in $X/G$; but that's forbidden
by $P$ being in $\fcb(X/G)$.  Thus, $A = I^-_G[c] \in
\fcb(X/G)$, and $\hpG(A) = P$.  

Consider how $\hpG$ works with the subset relation:  For
elements $A$ and $A'$ of $\fcb(X_G)$, generated
respectively by future chains $\{x_n\}$ and $\{x'_n\}$ in
$X_G$, we have $A \subset A' \iff$ for all $n$, there
is some $m$ with $x_n \ll_G x'_m \iff$ for all $n$, there
is some $m$ with $\pi(x_n) \ll \pi(x'_m) \iff \hpG(A)
\subset \hpG(A')$.  This also establishes the injectivity
of $\hpG$, restricted to $\fcb(X_G)$. \qed\enddemo

Proposition 2.6(2) tells us that our object of
interest---the future boundary of $X/G$---is entirely
reflected in the future boundary of the somewhat
mysterious chronological set $X_G$.  Then Proposition
2.3(d) tells us that that boundary is identifiable with
something we can get a handle on---the GIPs of $X$ making
up $\fcb_G(X)$.  (Corollary 2.5 shows us how to view GIPs
easily in term of IPs.)  

But what we really want out of all this is topology.

\subhead 2.2: Discussion of the chronological topology
\endsubhead \medskip

The topology of an intended boundary for a
spacetime---or, more generally, chronological set---is
addressed in \cite{H2}, where what is called the
\htop is defined for any chronological set; it may
perhaps be called the (future) chronological topology. 
This differs from the GKP topology on the causal boundary
(or even just the future causal boundary).  That is
intentional, as the GKP topology has some severe
problems; notably, the GKP topology for the causal
boundary of Minkowski space is not that of the conformal
embedding into the Einstein static spacetime, as one
would expect.  The \htop of
\cite{H2} does give the embedding topology for the future
causal boundary of Minkowski space, as well as for proper
embeddings of any strongly causal spacetime with a
spacelike causal boundary (see Theorem 5.3 of \cite{H2});
also, the \htop for the future causal boundary of any
standard static spacetime---metric product $\Bbb L^1
\times N$ for $N$ a Riemannian manifold---follows
directly from a geometric boundary construction on $N$
(see Theorem 6 of
\cite{H3}).  It also has such nice properties as making a
chronological set $X$ dense in $X \cup \partial(X)$ for
any sort of reasonable approximation of a future boundary
$\partial(X)$ (Theorem 2.4 of \cite{H2}).  Accordingly, it
is the \htop that will be used for boundaries here.

Note that in the case of a chronological action from a
group $G$ on a chronological set $X$, the group
automatically acts via homeomorphisms, using the \htop on
$X$:  This is because for any $g \in G$, the motion by
$g$---$\phi_g : X \to X$ defined by $\phi_g : x \mapsto g
\cdot x$---is an isomorphism of the chronology relation
(since $x \ll y$ implies $\phi_g(x) =
\phi_g(y)$ and $\phi_{g^{-1}} = (\phi_g)^{-1}$) and the
\htop is defined purely in terms of the chronology
relation.

The chief virtue of the \htop as used in \cite{H2} is the
fact (Proposition 2.7 in \cite{H2}) that in at least one
reasonable category, any map
$f: Z \to Y$ between chronological sets which is not only
future-continuous but also continuous in the \hto ies of
$Z$ and $Y$ yields a map $\hat f: \hat Z \to \hat Y$
between the future-completions which is again not only
future-continuous, but continuous in the \hto ies. 
However, the category in which this was proved is that of
chronological sets which are past-regular and
past-distinguishing and with spacelike future boundaries. 
Past-regularity is not a problem for our context of
$\pi_G : X_G \to X/G$ (as will be easily shown), and the
absence of past-distinguishing we've already dealt with in
establishing the existence of $\hpG$; but the issue of
spacelike boundaries is not readily sidestepped (and not
desirable to subsume).  Our tack must be to look at
continuity afresh, without specific reference to
Proposition 2.7 of \cite{H2}.

Past-regularity is needed in order to use a relatively
simple formulation for the \htop (a more complex
version being needed without that property being
present).  Past-regularity has to do with
whether IPs are what we naively expect:  A point
$x$ is called {\it past-regular\/} if $I^-(x)$ is an IP,
which is equivalent to there being a future chain $c$
such that $x$ is a future limit of $c$; a chronological
set is called past-regular if all its points are.  Thus,
any spacetime is past-regular (just take a future-directed
timelike curve $c$ terminating at $x$; $I^-(x) =
I^-[c]$, an IP), and if $X$ is a past-regular
chronological set, so is $\hat X$ (for any IP $P \in
\fcb(X)$, just take a future chain
$c$ generating $P$; $I_{\hat X}^-(P) = I_{\hat X}^-[c]$,
an IP of $\hat X$).  Thus, in our context of interest,
where $X$ and $X/G$ are both spacetimes, the only object
we need be at all concerned about is $X_G$; but this is
not a worry:

\proclaim{Proposition 2.7} Let $X$ be a chronological
set with a chronological action from a group
$G$.  If $X$ is past-regular, then so is $X/G$, and $X_G$
is past-regular if and only if $X/G$ is. \endproclaim 

\demo{Proof} Suppose $X$ is past-regular.  For any point
$p \in X/G$, pick a point $x \in \pi^{-1}(p)$; then there
is a future chain $c$ in $X$ with $I^-[c] = I^-(x)$. 
Then $\pi[c]$ is a future chain in $X/G$, and $I^-[\pi[c]]
= \pi[I^-[c]] = \pi[I^-(x)] = I^-(\pi(x)) = I^-(p)$, so
$p$ is past-regular.

Suppose $X_G$ is past-regular.  For any point $p \in
X/G$, pick a point $x \in \pi^{-1}(p)$; then there
is a future chain $c$ in $X_G$ with $I^-_G[c] =
I^-_G(x)$, \ie, $\cp G \cdot I^-[c] = \cp G \cdot
I^-(x)$; from this it follows that $\pi[I^-[c]] =
\pi[I^-(x)]$, \ie, $I^-[\pi[c]] = I^-(\pi(x)) = I^-(p)$. 
Since $\pi[c]$ is a future chain in $X/G$, this shows $p$
to be past-regular.

Suppose $X/G$ is past-regular.  For any point $x \in X$,
$\pi(x)$ is the future limit of some chain $\sigma
=\{p_n\}$ in $X/G$, \ie, $I^-[\sigma] = I^-(\pi(x))$. 
For each $n$, pick some $x_n \in \pi^{-1}(p_n)$; then $c
= \{x_n\}$ is a future chain in $X_G$.  We have
$\pi[I^-[c]] = I^-[\pi[c]] = I^-[\sigma] = I^-(\pi(x)) =
\pi[I^-(x)]$, from which it follows that $\cp G \cdot
I^-[c] = \cp G \cdot I^-(x)$, \ie, $I^-_G[c] = I^-_G(x)$;
thus, $x$ is past-regular in $X_G$. \qed \enddemo

(It is unclear whether $X/G$ can be past-regular without
$X$ also being past-regular; but this does not seem to be
of practical concern.)

For a past-regular chronological set $X$, the \htop is
defined thus:  Let $\ip(X)$ denote the set of IPs of
$X$.  For any sequence of points
$\sigma =
\{x_n\}$, define $L(\sigma)$ (the ``first-order" limits
of $\sigma$---plural, in case of non-Hausdorff contexts)
by $x \in L(\sigma)$ if and only if

\roster
\item for all $y \ll x$, eventually $y \ll x_n$ (\ie,
for $n$ sufficiently large, $y \ll x_n$) and
\item for any $P \in \ip(X)$ such that $P \supset I^-(x)$
but $P \neq I^-(x)$, there is some $y \in P$ such that
eventually $y \not\ll x_n$.
\endroster
$L$ is called the limit-operator for $X$.

We get a topology (the \hto y) by
defining a set $A$ to be closed if and only if for every
sequence $\sigma$ in $A$, $L(\sigma) \subset A$.  A
function $f : X \to Y$ between chronological sets will be
continuous in the \hto ies if for every sequence $\sigma$
in $X$, $f[L(\sigma)] \subset L(f[\sigma])$ (that is a
bit stronger than is needed, in strongly non-Hausdorff
cases---a necessary and sufficient condition involves
iteration of the set function $L[\;]$ an uncountable
number of times, up to the first uncountable
ordinal---but that is seldom of moment). Details are
in \cite{H2}. 

\proclaim{Proposition 2.8} Let $X$ be a
past-regular chronological set with a chronological
action from a group $G$. The map $\pi_G : X_G \to X/G$ is
continuous in the \hto ies, as is its extension $\hpG :
\hXG \to \hX/G$ to the future completions; furthermore,
for every sequence $\sigma$ and point $x$ in $\hXG$,
$\sigma$ converges to $x$ if $\hpG[\sigma]$ converges to
$\hpG(x)$. \endproclaim

\demo{Proof} Since $X$ is past-regular, by Proposition
2.7, so are $X_G$ and $X/G$, so we can use the definition
of \htop given by $L$ as above ($L$ will denote the
limit-operator in $X/G$, $L_G$ the one in $X_G$, with
$\hat L$ and $\hat L_G$ for $\hX/G$ and $\hXG$
respectively).

Note first that Lemma 2.2 and Proposition 2.3(c) combine
to yield a correspondence via $\pi$ between the IPs of
$X/G$ and the IPs of $X_G$ (which, by Corollary 2.5, are
the $G$-orbits of IPs of $X$); specifically, for $A \in
\ip(X_G)$, $\pi[A] \in \ip(X/G)$, and for $P \in
\ip(X/G)$, $\pi^{-1}[P] \in \ip(X_G)$.  Furthermore, this
correspondence clearly respects the subset relation:  For
$A \subset B$ IPs in $X_G$, $\pi[A]
\subset \pi[B]$, and for $P \subset Q$ IPs in $X/G$,
$\pi\inv[P] \subset \pi\inv[Q]$.  Thus, this
correspondence is a bijection between $\ip(X_G)$ and
$\ip(X/G)$.  Proposition 2.6(2) tells us that this
correspondence restricts (as $\bar\pi_G$) in perfect
measure to the future boundaries---and, consequently,
non-boundary IPs (\ie, pasts of points) correspond to
non-boundary IPs.

Let $\sigma = \{x_n\}$ be a sequence in $X$ and $x$ a
point in $X$.  First assume $x \in L_G(\sigma)$;
we want to show that $\pi(x) \in L(\pi[\sigma])$, \ie,
that it satisfies conditions (1) and (2) of the
definition of $L$.  Let $p = \pi(x)$. 
\roster 
\item 
For any $q \ll p$, pick $y \in \pi^{-1}(q)$; then $y
\ll_G x$.  Consequently, eventually $y \ll_G x_n$, which
implies $q \ll \pi(x_n)$.
\item 
For any $P \in \ip(X/G)$ with $P \supsetneq
I^-(p)$, we have $A = \pi\inv[P] \in \ip(X_G)$ with
$A \supsetneq \pi\inv[I^-(p)] = I^-_G(x)$. 
Consequently, there is some $y \in A$ with eventually $y
\not\ll_G x_n$, which implies $\pi(y) \not\ll \pi(x_n)$.
\endroster 
This establishes the continuity of $\pi_G: X_G \to X/G$
(since preservation of the $L$ operator implies
continuity).

Now assume $\pi(x) \in L(\pi[\sigma])$; we want to show
that $x \in L_G(\sigma)$.  (Again, $p = \pi(x)$.)
\roster 
\item 
For any $y \ll_G x$, we have $\pi(y) \ll p$, so
eventually $\pi(y) \ll \pi(x_n)$, which implies $y \ll_G
x_n$.  
\item 
For any $A \in \ip(X_G)$ with $A \supsetneq
I^-_G(x)$, we have $P = \pi[A] \in \ip(X/G)$ with $P
\supsetneq I^-(p)$.  Consequently, there is some $q \in
P$ with eventually $q \not\ll \pi(x_n)$.  Pick some $y
\in \pi\inv(q)$; then $y \in A$ and $y \not\ll_G x_n$.
\endroster
This establishes the odd sort of ``inverse-continuity" of
$\pi_G$ (by the same sort of inductive process which
establishes that preservation of $L$ is sufficient for
continuity, \ie, preservation of $L^\Omega$).

Now we expand to consideration of the future-completed
objects, $\hXG$ and $\hX/G$.  The considerations above
still suffice for sequences in $X$ and limits in $X_G$ or
in $X/G$, since there is very little additional matter
to be considered, and that little works out very
easily:  For instance, to show $\pi(x) \in
\hat L(\pi[\sigma])$ (with $\hat L$ denoting the
limit-operator in $\hX/G$, we must show additionally that
for any $P \in \fcb(X/G)$ with $P \ll p$, eventually $P
\ll \pi(x_n)$; but $P \ll p$ means there is some $q \ll
p$ with $P \subset I^-(p)$, and the rest follows.  Also,
for any past-regular chronological set $Y$,
$\ip(\widehat Y)$ and $\ip(Y)$, while not identical,
correspond very nicely:  For $P \in \ip(\widehat Y)$, $Y
\cap P \in \ip(Y)$, and for $Q \in \ip(Y)$, $\hat I^-[Q]
\in \ip(\widehat Y)$ (where $\hat I^-$ is the past in
$\widehat Y$), and this is a bijection respecting
$\subset$ (this is from  the Lemma in Theorem 2.4 of
\cite{H2}).  

Note that the correspondence of the chronology relations
between $X_G$ and $X/G$, used repeatedly in the arguments
above, extends to $\hXG$ and $\hX/G$ as well (see
Proposition 2.6(1)).  Thus the same arguments apply:

Let $\sigma = \{\alpha_n\}$ be a sequence in $\hXG$
and $\alpha$ an element of $\hXG$.  First assume $\alpha
\in \hat L_G(\sigma)$; we want to show $\hpG(\alpha) \in
\hat L(\hpG[\sigma])$.  Let $\theta = \hpG(\alpha)$.
\roster
\item 
For any $\phi \ll \theta$, there is some $q \in
X/G$ with $\phi \ll q \ll \theta$.  Pick $y \in
\pi\inv(q)$; then $y \ll_G \alpha$.  Consequently,
eventually $y \ll_G \alpha_n$, which implies $q \ll
\hpG(\alpha_n)$, hence $\phi \ll \hpG(\alpha_n)$.
\item 
For any $P \in \ip(\hX/G)$ with $P \supsetneq
\hat I^-(\theta)$, we have $A = \hpG\inv[P] \in \ip(\hXG)$
with $A \supsetneq \hpG\inv[\hat I^-(\theta)] = \hat
I^-_G(\alpha)$.  Consequently, there is some $\beta \in
A$ with eventually $\beta \not\ll_G \alpha_n$, which
implies $\hpG(\beta) \not\ll \hpG(\alpha_n)$.
\endroster
This establishes the continuity of $\hpG: \hXG \to \hX/G$.

Now assume $\hpG(\alpha) \in \hat L(\hpG[\sigma])$; we
want to show $\alpha \in \hat L_G(\sigma)$.  (Again,
$\theta = \hpG(\alpha)$.)
\roster
\item
For any $\beta \ll_G \alpha$, we have $\hpG(\beta) \ll
\theta$, so eventually $\hpG(\beta) \ll \hpG(\alpha_n)$,
which implies $\beta \ll_G \alpha_n$.
\item
For any $A \in \ip(\hXG)$ with $A \supsetneq \hat
I^-_G(\alpha)$, we have $P = \hpG[A] \in \ip(\hX/G)$ with
$P \supsetneq \hat I^-(\theta)$.  Consequently, there is
some $\phi \in P$ with eventually $\phi \not\ll
\hpG(\alpha_n)$.  There is some $q \in X/G$ with $\phi
\ll q \in P$; then $q \not\ll \hpG(\alpha_n)$.  Pick some
$y \in \pi\inv(q)$; then $y \in A$ and $y \not\ll_G
\alpha_n$.
\endroster
This establishes the inverse-continuity of $\hpG$.
\qed\enddemo

This completes the topological identification of the
future chronological boundary of $X/G$:

\proclaim{Theorem 2.9} Let $X$ be a past-regular
chronological set with a chronological action from a
group $G$.  Then $\fcb(X/G)$, the future chronological
boundary of the quotient space, is identifiable with
$\fcb_G(X)$, the future chronological $G$-boundary of the
total space, topologized as $\fcb(X_G)$, the future
chronological boundary of the $G$-expansion of the total
space.  

Specifically: The map $\pi^\partial_G : \fcb(X_G) \to
\fcb(X/G)$ is a homeomorphism in the \hto ies, and
$\fcb(X_G)$ is identified with $\fcb_G(X)$ via any IP in
$X_G$ being a GIP in $X$ and vice versa, with boundary
elements corresponding to one another.  The attachment of
$\fcb(X/G)$ to $X/G$ is exactly reflected in the
attachment of $\fcb_G(X)$ to
$X$---interpreted as  $\fcb(X_G)$ attaching to
$X_G$---via the map $\hpG: \hXG \to \hX/G$, in that a
sequence $\sigma$ in $\hXG$ converges to an element $A
\in \fcb(X_G)$ if and only if $\hpG[\sigma]$ converges to
$\hpG(A)$ in $X/G$. \endproclaim

\demo{Proof} The statement about sequences in $\hXG$ and
$\hpG$ comes directly from Proposition 2.8. 
Proposition 2.6 establishes that the restriction of
$\hpG$ to the boundaries is a bijection $\pi^\partial_G :
\fcb(X_G) \to \fcb(X/G)$; then the statements
about continuity and inverse-continuity of $\hpG$ amount
to saying that $\pi^\partial_G$ is a homeomorphism. 
Proposition 2.3 gives the identification of
$\fcb(X_G)$ with $\fcb_G(X)$. \qed \enddemo

\head 3: Quotients of Spacetimes with Spacelike Boundaries
\endhead

How do group actions on a spacetime extend to the
boundary, and what is the nature of the interaction
between the boundary and the group action?  The subject
of this section is to explore those questions, especially
in the context of spacelike boundaries.

The future chronological boundary of $X/G$, as a
topological space, may be only tenuously related to
$X$, the future chronological boundary of $X$, and the
group action, even in the simplest of settings. 

Consider, for instance, Minkowski 2-space $\mk2$; its
well-known future boundary $\fcb(\mk2)$
consists of two null lines ($\Cal I_L$ and $\Cal
I_R$, left and right future null infinity) joined at a
point ($i^+$ = future timelike infinity); see, e.g.,
\cite{HE}.  There is a simple action from the integers
$\Bbb Z$ on $\mk2$ given by $m \cdot (t,x) = (t,x+m)$. 
The quotient $\mk2/\Bbb Z$ is the Minkowski 2-cylinder,
the standard static spacetime $\mk1\times\Bbb S^1$ (with
a circumference of 1 for the circle $\Bbb S^1$).  By
Example 2 in \cite{H3}, the future chronological boundary
of the product spacetime $\mk1\times K$ is very simple
when $K$ is compact:  It is just a single point
$\{i^+\}$.  

The $\Bbb Z$-action on $\mk2$ extends
continuously (though not freely) to the boundary $\Cal
I_L \cup i^+ \cup \Cal I_R$.  We can readily parametrize
$\Cal I_L$, the IPs of $\mk2$ generated by null lines
rising to the left, as $\{P^L_a \,|\, a \in \Bbb R\}$ with
$P^L_a = \{(t,x) \,|\, t<-x+a\}$; similarly, $\Cal I_R = 
\{P^R_a \,|\, a \in \Bbb R\}$ with $P^R_a = \{(t,x) \,|\,
t<x+a\}$.  Then the homeomorphism $\phi_m : \mk2 \to
\mk2$ defined by $\phi_m : p \mapsto m \cdot p$ extends
continuously to $\fcb(\mk2)$ by $P^L_a \mapsto P^L_{a+m}$,
$P^R_a \mapsto P^R_{a-m}$, and $i^+ \mapsto i^+$ ($i^+$
being the IP consisting of all of $\mk2$).  So we can at
least consider the quotient $\fcb(\mk2)/\Bbb Z$; that
consists of two circles and a singular point (the
equivalence class of $i^+$) whose only neighborhood in
$\fcb(\mk2)/\Bbb Z$ is that entire space: a strongly
non-Hausdorff result.  This bears no apparent relation to
$\fcb(\mk2/\Bbb Z)$.

It is worth noting that there is another topology of
possible interest on $\fcb(\mk2)/\Bbb Z$, in that
$\widehat{\mk2}/\Bbb Z$ is a chronological set in
its own right, as in Proposition 2.1.  This holds
generally:  As Lemma 3.1 below will show, a chronological
action from a group $G$ on a past-regular chronological
set $X$ extends to a chronological $G$-action on
$\widehat X$, so that, by Propositions 2.1 and 2.7,
$\widehat X/G$ is also a past-regular chronological set;
it thus has its own \hto y.  Since the $G$-action
takes $X$ to $X$, the boundary $\fcb(X)$ is a
$G$-invariant subset of $\widehat X$, so $\fcb(X)/G$ is a
subspace of $\widehat X/G$ and inherits a subspace
topology.  

In the case of $\mk2$, the \hto y on
$\fcb(\mk2)/\Bbb Z$ (\ie, as a subspace of
($\widehat{\mk2}/\Bbb Z$, \hto y)), is totally
indiscrete:  For any two elements $P$ and $Q$ in
$\fcb(\mk2)$, with $\<P\>$ and $\<Q\>$ denoting the
equivalence classes under the $\Bbb Z$-action, $\<P\>
\in L(\sigma_{\<Q\>})$ with $\sigma_{\<Q\>}$ denoting the
constant sequence whose every element is $\<Q\>$.  To see
this, first note that for any $P \in \fcb(\mk2)$,
$I^-(\<P\>)$ is all of $\mk2/\Bbb Z$, since for any $p \in
\mk2$, there is some $n$ such that $p \in n \cdot P$, so
$p \ll n \cdot P$, so $\<p\>\ll \<P\>$.  It follows that
no IP of $\widehat{\mk2}/\Bbb Z$ properly contains
$I^-(\<P\>)$ (by Lemma 2.2 we know that the IPs of
$\widehat{\mk2}/\Bbb Z$ are the GIPs
of $\widehat{\mk2}$).  Thus, to see that $\<P\>
\in L(\sigma_{\<Q\>})$, we just need to note that for
every $\<p\> \ll \<P\>$, $\<p\> \ll \<Q\>$ also---true,
since $I^-(\<Q\>)$ is also $\mk2/\Bbb Z$.  In some sense
this wholly amorphous topological space is a better
reflection of $\fcb(\mk2/\Bbb Z)$ than that given by the
quotient topology:  It accurately suggests that we ought
to view all those elements of $\fcb(\mk2)/\Bbb Z$ as
really a single point.  But it is still not at all the
same thing as $\fcb(\mk2/\Bbb Z)$.

On the other hand, some spacetimes $V$ with group action
$G$ evince a very simple and natural relation between
$\fcb(V/G)$ and the $G$-action on $V$ extending to
$\fcb(V)$.  Consider, for example, ``lower" Minkowski
space, $\mk2_- = \{(t,x) \in \mk2 \,|\, t<0\}$, with the
same $\Bbb Z$-action as above.  The IPs of this spacetime
are easily calculated, so that $\fcb(\mk2_-)$ can be seen
to be $\{P_a \st a \in \Bbb R\}$ where $P_a = \{(t,x) \st
t < -|x-a| \}$; thus, $\fcb(\mk2_-) \cong \Bbb R$.  The
GIPs of $\mk2_-$ are also easily found (just $\cp \Bbb
Z \cdot P$ for any IP $P$), yielding the boundary-GIPs as
$\{B_a \st a \in \Bbb R\}$ with $B_a = \{(t,x) \st t <
-|x-a + m| \text{ for some } m \in \Bbb Z\}$; thus, $B_a
= B_{a+1}$, and by Theorem 2.9,  $\fcb(\mk2_-/\Bbb Z)
\cong \fcb_{\Bbb Z}(\mk2_-) \cong \Bbb S^1$, a circle.  

The $\Bbb Z$-action on $\mk2_-$ extends to the boundary by
$m \cdot P_a = P_{a+m}$, the typical action of the
integers on the real line.  The quotient by this action is
thus a circle; we have the simple result that
$\fcb(\mk2_-/\Bbb Z) \cong \fcb(\mk2_-)/\Bbb Z$.  (The
same topology arises from using \hto y on
$\widehat{\mk2_-}/\Bbb Z$.) It appears that the crucial
point is that both 
$\mk2_-$ and its quotient by $\Bbb Z$ have spacelike
boundaries---unlike $\mk2$, which has a null boundary.

The significance of spacelike boundaries is that the issue
of continuous extension of a map to the boundary is
considerably simplified if the boundary is spacelike.  In
brief:  Spacelike boundaries guarantee continuous
extension, but continuous extension can fail on timelike
boundaries (or even null boundaries---even in Minkowski
space), as is shown in \cite{H2}.  

The details run thus:  Call a point $x$ in a past-regular
chronological set $X$ {\it inobservable\/} if $I^-(x)$ is
not contained in any other IP; in effect, this says that
no observer observes all the events that constitute the
past of the event $x$, except for an observer who actually
experiences $x$.  Then $X$ is said to have {\it only
spacelike future boundaries\/} if all elements of
$\fcb(X)$ are inobservable, and if the set of all
inobservables is a closed subset of $\hat X$ in the
\htop.  If $X$ is a strongly causal spacetime $M$, the
latter condition follows from the first, since in a
spacetime proper, no points are inobservable---there's
always a point to the future---and with strong causality,
$\fcb(M)$ is necessarily closed in $\hat M$ (Proposition
2.6 in \cite{H2}).  A future-continuous map
$f: X \to Y$ is said to {\it preserve spacelike future
boundaries\/} if $\hat f :
\hat X \to \hat Y$ preserves inobservables (\ie, $\hat
f(p)$ is inobservable in $\hat Y$ if $p$ is inobservable
in $\hat X$); for $f : M \to N$ a map between strongly
causal spacetimes, both of which have only spacelike
future boundaries, this just amounts to saying that $\hat
f$ takes boundary points to boundary points.  Then
Proposition 2.7 of \cite {H2} says that if $f: X \to Y$
is a future-continuous map between past-regular
chronological sets having only spacelike future
boundaries, such that $f$ preserves spacelike boundaries
and is continuous in the \hto ies, then $\hat f: \hat X
\to \hat Y$ is also continuous in the \hto ies.  

For spacetimes:  If $M$ and $N$ are strongly causal
spacetimes, each satisfying the property that the entire
history of no endless observer is available to any other
observer, and if $f: M \to N$ is a continuous map
preserving the chronology relation such that for any
future-endless timelike curve $\sigma$ in $M$, $f \circ
\sigma$ is future-endless in $N$, then $\hat f : \widehat
M \to \widehat N$ is also continuous (in the \hto ies). 
If the hypotheses are not met, this goes wrong even in
simple cases:  \cite{H2} provides a counterexample with
$f: M \to \mk2$, where $M$ is all points to one side of a
timelike line in $\mk2$ ($f$ is continuous, but $\hat f$
is not); and this can be modified to $f' : M' \to \mk2$
with $M'$ being all points to one side of a null line
(map $M'$ to $M$, preserving chronology, by mapping null
lines onto themselves; then apply $f$).

It would thus seem that spacelike boundaries make for
simplified topological behavior when extending actions to
boundaries, and we may thus expect simple behavior in the
quotients, that the boundary of the quotient is the
quotient of the boundary.  With spacelike future
boundaries for a spacetime $V$ and its quotient $V/G$,
there is no difference between the quotient topology on
$\V/G$ and the \htop on that chronological set (see
Proposition 3.3).  Then the following seems to be likely:

\proclaim{Conjecture} Let $V$ be a spacetime
with a free, properly discontinuous, and chronological
action from a group $G$, with $M = V/G$ being strongly
causal, such that both $V$ and $M$ have only spacelike
future boundaries.  Then $\M$ is homeomorphic to
the quotient space $\V/G$, and $\fcb(M)$ is
homeomorphic to the quotient space $\fcb(V)/G$ (with \hto
ies used on $\M$ and $\V$); in other words,
$\widehat{V/G} \cong \V/G$ and $\fcb(V/G)
\cong \fcb(V)/G$.
\endproclaim

However, to obtain the conclusions in this conjecture,
one must also know that $\V/G$, as a chronological set,
is past-distinguishing (see Theorem 3.4).  This is a
rather unnatural kind of hypothesis, in that it does
not have any simple explication in terms of either of the
two spacetimes proper, $V$ and $M$; it would certainly be
desirable to do without it.  Is it a necessary
assumption, or does it follow from the other hypotheses? 
This is unclear at the present; no examples have come to
light as yet showing this assumption to be necessary.  We
will see that at least for a fairly largish class of
spacetimes---what are sometimes called multi-warped
spacetimes---it is only the spacelike boundaries that are,
indeed, the crucial ingredient, without need for an
explicit additional assumption on $\V/G$.

\medskip

The first step in approaching the study of these quotients
is to show that there is a group action on the boundary,
extending that on the spacetime.  This holds quite
generally, on chronological sets irrespective of the
nature of the boundary:

\proclaim{Lemma 3.1} Let $X$ be a
past-regular chronological set with a chronological
action from a group $G$. The $G$-action on $X$ extends
to a chronological $G$-action on $\widehat X$, continuous
in the \htop.
\endproclaim

\demo{Proof} For each $g \in G$. Let $\phi_g : X \to X$
be defined by $\phi_g : x \mapsto g \cdot x$.  In
order to be able to define $\hat\phi_g : \widehat X \to
\widehat X$, we need to show (just as in the proof of
Proposition 2.6 above) that for any future chain $c$ in
$X$ generating an element $P \in \fcb(X)$, $\phi_g[c]$
does not have two different future limits in $X$.  But
since $\phi_g$ is an isomorphism of the chronology,
relation, this is clear:  Since $c$ has no future limit
in $X$, neither can $\phi_g[c]$.  This allows the
definition of $\hat\phi_g$; for $P \in \fcb(X)$,
$\hat\phi_g(P) = g\cdot P$, so $\hat\phi_g \circ
\hat\phi_h = \hat\phi_{gh}$.  
  
Note that $\hat\phi_g$ is also an isomorphism of the
chronology relation:  If $ x \ll P$ for $x \in X$ and $P
\in \fcb(X)$, \ie, $x \in P$, then $g\cdot x
\in g \cdot P$, so $\hat\phi_g(x) \ll \hat\phi_g(P)$; if
$P \ll x$, \ie, $P \subset I^-(w)$ for some $w \ll x$,
then $g \cdot P \subset g \cdot I^-(w) = I^-(g\cdot w)$
and $g \cdot w \ll g\cdot x$, so $\hat\phi_g(P) \ll
\hat\phi_g(x)$; and if $P \ll Q$ ($Q \in \fcb(X)$), \ie,
$P \subset I^-(w)$ for some $w \in Q$, then $g\cdot P
\subset I^-(g\cdot w)$ and $g\cdot w \in g\cdot Q$, so
$\hat\phi_g(P) \ll \hat\phi_g(Q)$. It follows that
$\hat\phi_g$ is future-continuous and also a homeomorphism
in the \hto y. 

To show the action is chronological on $\widehat X$,
consider the possibility that $P \ll g \cdot P$ for $g
\neq e$.  This would mean that for some $w \in g \cdot P$,
$P \subset I^-(w)$.  Then $g^{-1}\cdot w \in P$, so
$g^{-1}\cdot w \in I^-(w)$, \ie, $g^{-1}\cdot w \ll w$;
but that violates the action of $G$ on $X$ being
chronological.
\qed\enddemo

It is worth noting that even in the simplest of
cases, the extension of a free $G$-action to the
boundary need not be free on the boundary.  The example
of $\mk2$ at the beginning of this section shows how the
free $\Bbb Z$ action on a spacetime can fail to be
free on the boundary ($i^+$ is a fix-point of the action
on $\widehat{\mk2}$).  For an example with
spacelike boundaries, consider a standard
static spacetime on a sphere: the metric product
$V = \mk 1 \times \Bbb S^n$.  Let $G$ be a typical free
group action on the sphere, such as
$\Bbb Z_2$ acting (in any dimension) via $p \mapsto -p$,
or $\Bbb Z_4$ acting on $\Bbb S^3$ as
multiplication by $i$ (\ie, $(x,y,z,w) \mapsto
(-y,x,-w,z)$); then $G$ acts chronologically (and freely
and properly discontinuously) on $V$ via
$g \cdot(t,p) = (t,g\cdot p)$.  The quotient $M = V/G$
is also a standard static spacetime, $M = \mk1 \times
(\Bbb S^n/G)$.  By the results in \cite{H3}, since the
Riemannian factors in both spacetimes are compact, the
future chronological boundaries for both $V$ and $M$ are
trivial, a single point: $i^+$, the entire spacetime;
clearly, $V$ and $M$ have only spacelike boundaries.  The
$G$ action on $\fcb(V) = \{i^+\}$ is the degenerate
action:  For all $g \in G$, $g \cdot i^+ = i^+$.

\medpagebreak

The second step is to show that the projection onto the
quotient defined by the group action extends to the
boundaries; this requires  strongly causal spacetimes
and, for continuity, spacelike boundaries.

\proclaim{Proposition 3.2} Let $V$ be a spacetime
with a free, properly discontinuous, and chronological
action from a group $G$, with $M = V/G$ being strongly
causal. 

a) The projection  $\pi: V \to M$ extends to
a future-continuous map $\hat\pi:
\V \to \M$, with $\hat\pi$ mapping $\fcb(V)$ onto
$\fcb(M)$.

b)  If $V$ and $M$ have only
spacelike future boundaries, then
$\hat\pi$ is continuous in the respective \hto ies.
\endproclaim

\demo{Proof} As in the proof of Proposition 2.6, with no
assumptions being made of past-distinguishing or
past-determination, we need to show that for any future
chain
$c$ in $V$, $\pi[c]$ cannot have two future
limits in $M$, in order that $\hat\pi$ can be defined;
and, once it is known to exist, we must check separately
that $\hat\pi$ preserves the chronology relation.  

By Proposition 2.2 of \cite{H2}, a future limit of a
future chain is the same as a limit of the sequence in
the \hto y; and by Theorem 2.3 of \cite{H2}, for a
strongly causal spacetime the \htop is the same as the
manifold topology.  Thus, what's needed for the existence
of $\hat\pi$ is to show that for any future-endless
timelike curve $c$ in $V$, $\pi\circ c$ cannot have two
future endpoints in $M$; but this is just a consequence
of the manifold topology being Hausdorff.  

In fact, if $c$ generates an element of $\fcb(V)$,
$\pi\circ c$ cannot have any limit in $M$ (this will be
useful for part (b)).  For suppose
$\pi\circ c$ has $p \in M$ as a limit.  Let $U$ be a
neighborhood of $p$ such that $\pi^{-1}[U]$ is a set of
components $\{W_g \,|\, g \in G\}$ evenly covering $U$,
\ie, $\pi$ is a homeomorphism from each $W_g$ to $U$; let
$x_g$ be the pre-image of $p$ in $W_g$.  For
$U$ sufficiently small, $\pi\circ c$ enters $U$ and does
not leave it; consequently, $c$ cannot exit any $W_g$, so
it enters exactly one $W_h$ and never leaves it.  Then,
since $\pi: W_h \to U$ is a homeomorphism, $c$ must have
$x_h$ as a limit, violating the fact that $c$ is
future-endless.  

This allows for the definition of $\hat\pi : \widehat V
\to \widehat M$ in such a way that boundary points are
mapped to boundary points:  For a future-endless curve
$c$ generating $P \in \fcb(V)$, $\hat\pi(P) =
I^-[\pi\circ c] = \{\pi(x) \st  x \ll_G c(t)
\text{ for some } t \}  = \{\pi(x) \st  x \ll c(t) \text{
for some } t \} = \pi[P]$.  As in Proposition 2.6, to
show $\hat\pi$ is future-continuous, we need only show
that it is chronological:

Consider $x, y \in V$ and $P,Q \in \fcb(V)$.  If $x \ll
y$, then $\hat\pi(x) = \pi(x) \ll \pi(y) = \hat\pi(y)$. 
If $x \ll P$, \ie, $x \in P$, then $\hat\pi(x) = \pi(x)
\in \pi[P] = \hat\pi(P)$, \ie, $\hat\pi(x) \ll
\hat\pi(P)$.  If $P \ll y$, \ie, $P \subset I^-(w)$ for
some $w \ll y$, then $\hat\pi(P) = \pi[P] \subset
I^-(\pi(w))$ and $\pi(w) \ll \pi(y) = \hat\pi(y)$, \ie,
$\hat\pi(P) \ll \hat\pi(y)$.  If $P \ll Q$, \ie, $P
\subset I^-(w)$ for some $w \in Q$, then $\hat\pi(P) =
\pi[P] \subset I^-(\pi(w))$ and $\pi(w) \in \pi[Q] =
\hat\pi(Q)$, \ie, $\hat\pi(P) \ll \hat\pi(Q)$.  Thus,
$\hat\pi$ is chronological, and part (a) is established.

For part (b), we will use Proposition 2.7 of \cite{H2}. We
 are now assuming that both $V$ and $M$ have only
spacelike future boundaries; since they are manifolds,
they are automatically past-regular.  By Proposition 2.1,
$\pi : V \to M$ is future-continuous.  Then the only
remaining fact we need is that $\hat\pi: \V \to \M$
preserve inobservables; then Proposition 2.7 of \cite{H2}
will tell us that $\hat\pi$ is continuous in the \hto
ies.  

Since $V$ is a
spacetime, the only inobservables in $\V$ are the
boundary elements (all of which are inobservable, since
$V$ has only spacelike future boundaries).  If
$P \in \fcb(V)$ is generated by the future-endless
timelike curve $c$, then $\hat\pi(P)$ is generated by
$\pi\circ c$.  All we need is that $\hat\pi(P)$ is
inobservable, \ie, not contained in any other IP.  We
know $\hat\pi(P) = I^-[\pi\circ c]$.  Since $c$ is
future-endless, so is $\pi\circ c$ (as shown above). 
That means there is no $q \in M$ so that $I^-(q) =
\pi[P] = \hat\pi(P)$; in other words, $\hat\pi(P) \in
\fcb(M)$.  But since $M$ is assumed to have only
spacelike future boundaries, all elements of $\fcb(M)$ are
inobservable. \qed\enddemo

The hypotheses are really needed for Proposition 3.2. 
As an example of what can go wrong for part (a) if $M$ is
not strongly causal, let $V = \mk2$, and let the integers
$\Bbb Z$ act on $ \mk2$ in a null fashion:
$n \cdot (x,t) = (x+n,t+n)$.  Note that the quotient
$M = \mk2/\Bbb Z$ is not past-distinguishing (as well as
not strongly causal):  Let $[x,t]$ denote the equivalence
class of $(x,t)$; then $I^-([x,t]) = \{[y,s] \st s-y <
t-x\}$, so that $I^-([x,t]) = I^-([x+a,t+a])$ for all $a$,
a circle's worth of points in $M$, all with the same
past.   Consider the boundary IP $P = \{(x,t) \st t<x\}$;
it is generated by the future chain $c(n) =
(n,n-\frac1{n})$.  Examine $\pi[c] =
\{[n,n-\frac1{n}]\} = \{[0,-\frac1{n}]\}$: 
$I^-[\pi[c]]$ is not a boundary IP, since it is
the same as $I^-([0,0])$; but that is the same as any
$I^-([a,a])$.  Thus, there is no way to define
$\hat\pi(P)$, as it is supposed to be each $[a,a]$
(\ie, each $[a,a]$ is a future limit of $\pi[c]$).

An example of what can go wrong for part (b) if the
spacetimes have timelike boundaries will be given at the
end of this section (it is rather complicated but worth
knowing about).

\medpagebreak

As was mentioned before, we really have two topologies to
consider for $\V/G$ (hence, for $\fcb(V)/G$):  With the
\htop on $\V$, we can consider the quotient topology on
$\V/G$.  Alternatively, since $\V$ is a past-regular
chronological set with a chronological $G$-action, $\V/G$
is also (by Proposition 2.7) a past-regular chronological
set, yielding a \htop.  As the example of $\mk2$ at the
beginning of this section showed, these topologies need
not be the same in general; but with spacelike
boundaries, they are.  

\proclaim{Proposition 3.3} Let $V$ be a spacetime
with a free, properly discontinuous, and chronological
action from a group $G$, with $M = V/G$ being strongly
causal.   Suppose both $V$ and $M$ have only
spacelike future boundaries; then the quotient topology
and \htop on $\V/G$ are the identical, \ie, $(\V/G,
\text{\rom{quotient topology}}) \cong  (\V/G,
\text{\rom{\htop}})$. 
\endproclaim

\demo{Proof} For any topological space $A$ with a
$G$-action, the quotient map $\pi_A : A \to A/G$ is
continuous (with the quotient topology on $A/G$) and has
the following universal property:  If
$f : A \to B$ is any continuous $G$-invariant map, then
there is a unique continuous map $f_G : A/G \to B$ such
that $f_G \circ \pi_A = f$;  $f_G$ is defined
by $f_G: \<a\> \mapsto f(a)$, where $\<a\>$ denotes the
$G$-equivalence class of $a$.

Since $V$ is past-regular, so is $\V$, hence also
$\V/G$ (Proposition 2.7).  Consider the quotient map $\Pi
: \V \to \V/G$ as a function between chronological sets. 
Note that this is not the map $\hat\pi = \widehat{\pi_V}$
from Proposition 3.2.  On the set-level, this is the same
as the universal map $\pi_{\V}$; but we will use $\Pi$
instead of $\pi_{\V}$ to denote that we want to use the
\htop on the target space, which may ({\it a priori\/})
be different from the quotient topology there.  

We need to see that $\Pi$ is continuous.  We know
that $V/G$ is an open subset of $\V/G$ and that on that
open set, the \htop from $\V/G$ is the same as the \htop
on $V/G$ as a chronological set in its own right (Theorem
2.4 in \cite{H2}); and we know that that is the same as
the manifold topology on $V/G = M$.  We also know that
$\Pi$ takes boundary points to boundary points, so
that it restricts to $\pi_V : V \to M$.  That map is
continuous; thus, all we need be concerned about is the
boundary in $\V$.

So consider a sequence of boundary elements
$\sigma = \{P_n\}$ in $\V$.  The boundary is closed, so
the only possible members of $L(\sigma)$ are boundary
elements also.  Suppose $Q \in L(\sigma)$; we need to
show that $\Pi(Q) \in L(\Pi[\sigma])$.  Let $\<\;\>$
denote equivalence class under the $G$-action.  Since
$V/G$ has only spacelike future boundaries, all we need do
is consider if for all $\<x\>
\ll \Pi(Q) = \<Q\>$, eventually
$\<x\> \ll \Pi(P_n) = \<P_n\>$.  But $\<x\> \ll
\<Q\>$ if and only if $x \ll_G Q$, \ie, if for some $g
\in G$, $g \cdot x \in Q$.  If this obtains, then, as $Q
\in L(\{P_n\})$, eventually $g \cdot x \in P_n$; and that
yields $\<x\> \ll \<P_n\>$.  Hence, $\Pi$ is
continuous.

With $\Pi: \V \to (\V/G,\text{\rom{\htop}})$ continuous,
the universal property of $\pi_{\V}$ yields for us the
continuous map $\Pi_G : (\V/G, \text{\rom{quotient
topology}}) \to (\V/G,\text{\rom{\htop}})$; since
$\Pi$ is just $\pi_{\V}$ on the set-level, $\Pi_G$ is
just the identity map on the set-level.  We need to
show this map is proper, \ie, that limits in the target
space correspond to limits in the domain.  Since both
topologies yield the manifold topology on
$V/G = M$, we need only look at the boundary-points
for limits.     

So consider $Q \in \fcb(V)$ and a sequence $\sigma =
\{\alpha_n\}$ in $\V$ (each $\alpha_n$ either in $V$ or in
$\fcb(V)$, it doesn't matter) with $\<Q\> \in
L(\{\<\alpha_n\>\})$; we need to see that $\<Q\>$ is a
limit point of $\<\sigma\> = \pi_V[\sigma]$ in the
quotient topology.  Let $Q$ be generated by a future
chain $\{y_m\}$.  For all $m$, $\<y_m\> \ll \<Q\>$, so
$\<y_m\> \ll \<\alpha_n\>$ for $n$ sufficiently large,
i.e, there is some number $k_m$ and for all $n \ge k_m$
there is some $g^n_m \in G$ such that
$y_m \ll g^n_m \cdot \alpha_n$.  Since $\{y_m\}$ is a
future chain, this means that for all $m$, for all $i \le
m$, for all $n \ge k_m$, $y_i \ll g^n_m\cdot \alpha_n$. 
We can turn this around:  For all $i$, for all $m \ge i$,
for all $n \ge k_m$, $y_i \ll g^n_m \cdot \alpha_n$; in
particular (letting $n = k_m$), for all $i$, for all $m
\ge i$, $y_i \ll g^{k_m}_m \cdot \alpha_{k_m}$.  Let $\bar
\alpha_m =  g^{k_m}_m \cdot \alpha_{k_m}$; then we've
just shown that for all $i$, $y_i \ll \bar \alpha_m$ for
all $m \ge i$.  Since $V$ has only spacelike future
boundaries, this is sufficient to show that $Q \in
L(\bar\sigma)$, where $\bar\sigma = \{\bar \alpha_m\}$. 
Then, since $\pi_{\V}$ is continuous, $\<Q\>$ is a limit
of the sequence $\<\bar\sigma\> = \pi_{\V}[\bar\sigma]$.

That's not quite what we wanted; we need $\<Q\>$ a limit
of $\<\sigma\>$.  But it's actually good enough: 
For if we consider any subsequence $\tau \subset \sigma$,
then the same procedure will produce a subsubsequence
$\bar\tau \subset \tau$ with $\<Q\>$ a limit point
of $\<\bar\tau\>$.  With $\<Q\>$ a limit point of
some subsubsequence of every subsequence of
$\<\sigma\>$, we conclude $\<Q\>$ is a limit of
$\<\sigma\>$. \qed \enddemo

The last step is to show that the projection from $\V$
to $\M$ induces a homeomorphism between the quotient space
$\V/G$ and $\M$; this requires all the hypotheses of
Proposition 3.2(b) and, it seems, one additional
hypothesis.  It is easy enough to show there is a
continuous map between these spaces; the trick is to show
it is injective.  This can be accomplished by requiring
that $\V/G$ be past-distinguishing; but it is unclear
whether this is a truly necessary hypothesis.

\proclaim{Theorem 3.4} Let $V$ be a spacetime
with a free, properly discontinuous, and chronological
action from a group $G$, with $M = V/G$ being strongly
causal, such that both $V$ and $M$ have only spacelike
future boundaries.  Suppose that $\V/G$ is
past-distinguishing.  Then
$\M$ is homeomorphic to the quotient space $\V/G$, and
$\fcb(M)$ is homeomorphic to $\fcb(V)/G$ (with \hto ies
used on $\M$ and $\V$).
\endproclaim

\demo{Proof} Let $\pi : V \to M$ be projection.  By
Proposition 3.2, $\pi$ extends to the future-completions
$\hat\pi : \V \to \M$, and this is continuous in the
respective \hto ies.   Clearly, $\pi$ is $G$-invariant,
so $\hat\pi$ is also $G$-invariant when restricted to
$V$.  As shown in the proof of
Proposition 3.2, the action of $\hat\pi$ on boundary
elements is simple for $P \in \fcb(V)$: $\hat\pi(P) =
\pi[P]$.  Thus, $\hat\pi$ is also $G$-invariant on
$\fcb(V)$: $\hat\pi(g \cdot P) = \pi[g\cdot P] = \pi[P] =
\hat\pi(P)$.  It follows that there is
an induced map $\piG : \V/G \to \M$ (so that with
$\pi_{\V}: \V \to \V/G$ as in the proof of Proposition
3.3, $\piG \circ \pi_{\V} = \hat\pi$), which is
continuous using the quotient topology on $\V/G$.  By
virtue of Proposition 3.3, that is the same as the \htop
on $\V/G$.

Note that the $G$-action on $\V$ preserves $\fcb(V)$: For
$x\in V$, $g\cdot x \in V$ also.  Thus, $\V/G$ can be
broken into two parts: $\V/G = [V/G] \cup [\fcb(V)/G] = M
\cup [\fcb(V)/G]$.  Note that $\hat\pi : V \cup \fcb(V)
\to M \cup \fcb(M)$ neatly splits as $\pi : V \to M$ and a
map we can call $\pi^\partial : \fcb(V) \to \fcb(M)$
(this was contained in the proof of Proposition
3.2(b)).  It follows that $\piG : \V/G \to \M$ splits as
$\text{id}_M : M \to M$ and a map
$\piGb : \fcb(V)/G \to \fcb(M)$.

Note that $\text{id}_M$
really is the identity on $M$ in the topological
category:  By Theorem 2.4 of \cite{H2}, the topology on a
spacetime $N$ induced, as a topological subspace, by the
\htop on $\widehat N$ is the same as the \htop on $N$ as a
chronological set; that latter topology, by Theorem 2.3
of \cite{H2}, is the same as the manifold topology if $N$
is strongly causal; and (again, for $N$ being strongly
causal) by Proposition 2.6 of \cite{H2}, $\fcb(N)$ is
closed in $\widehat N$ in the
\htop.  It follows that the \htop on $\V$ induces the
manifold topology on $V$ (note that Proposition 1.3 tells
us $V$ is strongly causal), and that the quotient topology
on $\V/G$ induces the manifold topology on the open subset
$V/G = M$. 

So we need to concentrate on the nature of $\piGb :
\fcb(V)/G \to \fcb(M)$, which is defined thus:  Let
$P$ be in $\fcb(V)$, and let $\<P\>$ denote the
equivalence class of
$P$ under the $G$-action on $\fcb(V)$ (and also
use $\<x\>$ for the equivalence class of $x \in V$); then
$\piGb(\<P\>) = \pi^\partial(P) = \pi[P]$.  The most
important aspect to note is that this is injective (that
is what typically fails to be the case with non-spacelike
boundaries).  Injectivity of $\piGb$ comes directly from
the assumption that $\V/G$ is past-distinguishing:  

Consider $P \in \fcb(V)$ and its image $\<P\>$ in
$\V/G$.  What is the past of $\<P\>$ in that
chronological set?  For $x \in V$, $\<x\> \ll \<P\>$ if
and only if in $\V$, $x \ll_G P$, \ie, for some $g \in G$,
$x \ll g \cdot P$, \ie, for some $g$, $x \in g \cdot P$,
\ie, $x \in \bigcup \<P\>$ (thinking of $\<P\>$ as a
collection of subsets of $V$); and that last is equivalent
to  $\<x\>$ being among $\{\<y\> = \pi(y) \st y \in P\}$. 
It follows that $\<x\> \ll \<P\>$ if and only if $\<x\>
\in \pi[P] = \piGb(\<P\>)$.  What about other elements of
$\V/G$ possibly being in the past of $\<P\>$?  For $Q \in
\fcb(V)$, $\<Q\> \ll \<P\>$ if and only if for some $g
\in G$, $Q \ll g \cdot P$, \ie, for some $w \in g\cdot
P$, $Q \subset I^-(w)$; but that is impossible since $V$
has only spacelike future boundaries.  Therefore, we can
conclude that $\piGb(\<P\>) = I^-(\<P\>)$; thus,
$\piGb$ is injective precisely when $\V/G$ is
past-distinguishing. 

So we now know that $\piG : \V/G \to \M$ is a continuous
bijection.  By Proposition 3.3, we know that the quotient
topology on $\V/G$ is the same as the \htop using the
chronological structure on $\V/G$.  Thus, we can show
$\piG$ is a homeomorphism by showing that it is an
isomorphism of chronological sets.  First consider $x$
and $y$ in $V$:  There is no distinction between their
images in $\V/G$ and in $\M$ (\ie, $\<x\> = \pi(x)$) and
the chronology relation in either case resolves to $x
\ll_G y$.  Next consider also $P$ and $Q$ in $\fcb(V)$. 
As shown above, we have $\<x\> \ll \<P\>$ if and only if
$\<x\> \in \pi[P]$; but $\<x\> = \piG(\<x\>)$ and $\pi[P]
= \piG(\<P\>)$, so this is equivalent to $\piG(\<x\>) \in
\piG(\<P\>)$, and in $\M$ that is the same as $\piG(\<x\>) 
\ll \piG(\<P\>)$.  Finally, we cannot have either $\<P\>
\ll \<x\>$ or $\<P\> \ll \<Q\>$, as that would mean $P
\ll_G x$ and $P \ll_G Q$ respectively, resulting in $P
\ll g \cdot x$ and $P \ll g \cdot Q$, impossible for $V$
having only spacelike future boundaries; and as $M$ has
only spacelike future boundaries, we likewise cannot have
$\piG(\<P\>) \ll \piG(\<x\>)$ (\ie, $\pi[P] \ll \pi(x)$)
or $\piG(\<P\>) \ll \piG(\<Q\>)$ (\ie, $\pi[P] \ll
\pi[Q]$).  Therefore, in all cases, $\piG(\alpha) \ll
\piG(\beta)$ if and only if $\alpha \ll \beta$.

Since $\piG: \V/G \to \M = \widehat{V/G}$
exactly duplicates the chronology relations, it is a
homeomorphism.  The same is true for the restriction to
the closed subsets $\piGb: \fcb(V)/G \to \fcb(M) =
\fcb(V/G)$. \qed \enddemo

\subhead
Example: multi-warped spacetimes
\endsubhead

It is unclear whether the hypothesis in Theorem 3.4 that
$\V/G$ be past-distin\-guishing is actually necessary;
it seems possible that it may be a consequence of the
other hypotheses.  Here is a class of spacetimes in which
that is true: multi-warped spacetimes (called multiply
warped in \cite{H2}, but referred to as ``multi-warped"
in some recent publications, such as \cite{GS}).  These
are spacetimes which have the form $(a,b) \times_{f_1} K_1
 \cdots \times_{f_m} K_m$, where $(a,b)$ is an
interval, finite or infinite, in $\mk1$, each $K_i$ is a
Riemannian manifold with metric $h_i$, each $f_i : (a,b)
\to \Bbb R$ is a positive function, and the spacetime
metric is $-(dt)^2 + f_1(t)h_1 + \cdots + f_m(t)h_m$. 
These include inner Schwarzschild, Robertson-Walker
spaces, the Kasner spacetimes, and many others.
 
It was shown in \cite{H2} (Proposition 5.2) that if $M$ is
a multi-warped spacetime such that for each $i$, 
$(K_i,h_i)$ is complete and for some finite $b^- < b$,
$\int_{b^-}^b f_i^{-1/2} < \infty$, then $\fcb(M)$ is
spacelike and is homeomorphic to $K = K_1 \times \cdots
\times K_m$, with $\M$ homeomorphic to $(a,b] \times K$
and $\fcb(M)$ appearing there as $\{b\} \times K$. 
Before examining the behavior of multi-warped spacetimes
for Theorem 3.4, it is perhaps well to expand the
conditions under which it can be known that $\fcb(M)$ is
spacelike.  This is somewhat lengthy, but fairly
straight-forward.

\proclaim{Proposition 3.5}  Let $M = (a,b) \times_{f_1} 
K_1 \cdots \times_{f_m} K_m$ be a multi-warped
spacetime.  Let the Riemannian factors be arranged
such that (for some finite $b^-<b$) the first $k$ warping
functions---$i$ from 1 to $k$---obey $\int_{b^-}^b
f_i^{-1/2} < \infty$, and the rest---$i$ from $k+1$ to
$m$---obey $\int_{b^-}^b f_i^{-1/2} = \infty$.  Then the
following hold:
\roster
\item "(a)" If some Riemannian factor $K_i$ is incomplete,
then
$\fcb(M)$ has timelike-related elements.
\item "(b)" If for some $i \ge k+1$,  $K_i$ is not
compact, then $\fcb(M)$ has null-related elements, \ie,
$P \subset P'$ with $P \neq P'$.
\item "(c)" If neither of those occur, then $M$ has only
spacelike future boundaries.
\endroster

In the last case, $\fcb(M)$ is
homeomorphic to $K^0 = K_1 \times \cdots \times K_k$. 
Furthermore, $\M$ is homeomorphic to $\left((a,b] \times
K^0 \times K'\right)/\sim\,$, where $K' = K_{k+1} \times
\cdots \times K_m$ and $\sim$ is the equivalence relation
defined by $(b,x^0,x') \sim (b,x^0,y')$ for any $x^0 \in
K^0$ and $x',y' \in K'$; $\fcb(M)$ appears there as
$\{b\} \times K^0 \times \{*\}$.
\endproclaim 

\demo{Proof}  For part (a), suppose $K_i$ is incomplete.
Let $M_i = (a,b) \times_{f_i}  K_i$; then $M_i$ is
conformal to the product static spacetime
$\bar M_i = (a',b') \times K_i$, where $b' = d +
\int_d^b f_i^{-1/2}$ and $a' = d -
\int_a^d f_i^{-1/2}$ for some choice of $d$ between
$a$ and $b$.  It is easy to find timelike related elements
$P_i \ll P'_i$ in $\fcb(\bar M_i)$, such as by using
an endless unit-speed geodesic $c_i$ of finite length in
$K_i$ and letting $P_i$ and $P'_i$ be generated,
respectively,  by the null geodesics
$\sigma_i(t) = (t_0 + t, c_i(t))$ and $\sigma'_i(t) =
(t'_0  + t, c_i(t))$  with $t'_0 > t_0$; the numbers
$t_0$ and $t'_0$ have to be chosen so that, with
whatever the domain of $c_i$ is, $\sigma_i$ and
$\sigma'_i$ both remain within the comprehension of
$\bar M_i$, but this is not difficult.  We may
consider $\sigma_i$ and $\sigma'_i$ to be curves in
$M_i$ as well as in $\bar M_i$, though their
first coordinates will be parametrized differently:
$\sigma_i(t) = (\tau(t), c_i(t))$ and $\sigma'_i(t) =
(\tau'(t), c_i(t))$.  Note that although
$\sigma_i$ and $\sigma'_i$ are likely not to be geodesics
in $M_i$, they are still null curves in
$M_i$; the sets $P_i$ and $P'_i$ are still IPs in
$M_i$ generated by $\sigma_i$ and $\sigma'_i$, they are
elements in the future chronological boundary of $M_i$,
and they still obey $P_i \ll P'_i$ in $\M_i$:  There is
some $(s,x_i) \in M_i$ such that $P_i \subset
I^-((s,x_i))$ and $(s,x_i) \in P'_i$.

Now pick points $x_j \in K_j$ for each $j \neq i$, and
let $c$ be the curve in $K = K_1 \times \cdots
\times K_m$ defined by the projection to $K_i$ being
$c_i$ and the projection to each
$K_j$ being constant at $x_j$ for $j \neq i$.  Let
$\sigma$ and $\sigma'$ be curves in $M$ defined by
$\sigma(t) = (\tau(t), c(t))$ and $\sigma'(t) =
(\tau'(t),c(t))$; these are clearly endless null curves. 
Let $P$ and $P'$ be the IPs in $M$ generated by $\sigma$
and $\sigma'$ respectively; they are boundary elements. 
Let $x = (x_1,\dots,x_m)$ (so that this is $x_i$ in the
projection to $K_i$).  Note that as the past in $M_i$ of
$(s+\delta,x_i)$ (for $\delta > 0$ but small) includes
$\sigma_i$, it follows that the past in $M$ of
$(s+\delta,x)$ includes $\sigma$; hence, $P \subset
I^-((s+\delta,x))$.  Similarly, as $(s+\delta,x_i)$ is in
the past of $\sigma'_i$, it follows that $(s+\delta,x)$ is
in the past of $\sigma'$, \ie, $(s+\delta,x) \in P'$. 
Therefore, $P \ll P'$.

For part (b), suppose $K_i$ is not compact and
$\int_{b^-}^b f_i^{-1/2} = \infty$.  If $K_i$ is not
complete, then we are done, by part (a); so we may assume
$K_i$ is complete.  Any complete, non-compact Riemannian
manifold contains a ray---a geodesic minimizing over all
intervals---of infinite length (consider any sequence of
points $\{p_n\}$ with distances from $p_0$ going to
infinity and minimizing unit-speed geodesics $\gamma_n$
from $p_0$ to $p_n$; the tangent vectors
$\{\dot\gamma_n(0)\}$ have an accumulation vector $v$,
and that generates the requisite ray).  Let $c_i$ be
this ray in $K_i$.  As in part (a), let $M_i = (a,b)
\times_{f_i} K_i$ and let $\bar M_i$ be the
conformally related product static spacetime $(a',b')
\times K_i$; but in this case, we know $b' = \infty$. 
As before, let $\sigma_i$ and $\sigma'_i$ be the null
geodesics $\sigma_i(t) = (t_0, c_i(t))$ and $\sigma'_i(t)
= (t'_0, c_i(t))$ with $t'_0 > t_0$, generating IPs
$P_i$ and $P'_i$ respectively.  The results of \cite{H3}
on the future chronological boundary of standard static
spacetimes apply in this case, even if $a'$ is
finite:  The Busemann function for $c_i$ is finite,
since $c_i$ is a ray; thus, $P_i$ and $P'_i$ are distinct
elements in $\fcb(M_i)$, with $P_i \subset P'_i$.  (The
material in \cite{H3} shows that elements of $\fcb(\mk1
\times N)$ are generated by functions of the form $b_c(p)
= \lim_{t \to \infty}(t-d(p,c(t)))$ for $c$ a curve in
$N$ of no more than unit speed; that is the Busemann
function for $c$.)

We proceed as before, selecting points $x_j \in
K_j$ for $j \neq i$ and using them to define
$\sigma$ and $\sigma'$ as endless null curves in $M$,
generating boundary IPs $P$ and $P'$.  As $\sigma_i$ lies
in the past of $\sigma_i$, so does $\sigma$ like in the
past of $\sigma'$; hence, $P \subset P'$.  As there exists
some point $(s,x_i)$ in $P'_i$ but not in $P_i$, so does
$(s,x)$ lie in $P'$ but not in $P$, where $x$ has
projections to $x_j$ in $K_j$. 

For part (c), we emulate the proof for the case $k = m$
given in Proposition 5.2 of \cite{H2}.  First let
$\bar M = \left((a,b] \times K^0 \times K'\right)/
\sim\,$, with $(b,x^0,x') \sim (b,x^0,y')$ for any $x^0
\in K^0$ and any $x',y' \in K'$.  Let $[x^0]$ denote the
equivalence class of $(b,x^0,x')$.  Define a chronology
relation on $\bar M$, extending that on $M$, as follows: 
For $(t,x^0,x') \in M$ and $y^0 \in K^0$, set $(t,x^0,x')
\ll [y^0]$ if and only if there is a timelike curve in
$M^0 = (a,b) \times_{f_1} K_1 \cdots \times_{f_k} K_k$
starting at $(t,x^0)$ and approaching $(b,y^0)$.  This
amounts to saying there is a continuous curve $c^0: [t,b]
\to K^0$, going from $x^0$ to $y^0$, differentiable on
$[t,b)$, with $\sum_{i = 1}^k f_i(s)^{-1/2}
|\dot c^0_i|_i < 1$ for all $s \in [t,b)$, where $c^0_i$
denotes the projection of $c^0$ to $K_i$ and $|\;|_i$ is
the norm in the Riemannian metric on $K_i$.  Let $\bar
I^+$ and $\bar I^-$ denote the future and past operators
in the chronological set $(\bar M, \ll)$.

This amounts to the same construction used in Proposition
5.2 of \cite{H2}, applied to the spacetime $M^0$, save
that in that case, the chronological relation was defined
on $\bar M^0 = (a,b] \times K^0$ instead of
$\bar M = \left((a,b) \times K^0 \times
K'\right)/\sim\,$.  More precisely:  $(t,x^0,x') \ll
[y^0]$ in $\bar M$ if and only if $(t,x^0) \ll (b,y^0)$
in $\bar M^0$.  The Lemma of Proposition 5.2 assures us
that for any $(t,x^0) \in M^0$, $\bar I^+((t,x^0))$ is
open in $\bar M^0$, \ie, the set of $y^0 \in K^0$ such
that $(b,y^0) \gg (t,x^0)$ is open.  It follows that $\bar
I^+(t,x^0,x')$ is open in $\bar M$.

For any $x^0 \in K^0$, define $Q_{x^0} \subset M$
to be $\bar I^-([x_0])$.  The analogous set in $M^0$ is
$Q^0_{x^0}$, the past of $(b,x^0)$; evidently, $Q_{x^0}
= Q^0_{x^0} \times K'$.  We will see that
$\fcb(M) = \{Q_{x^0} \st x^0 \in K^0\}$, just as in
Proposition 5.2 of \cite{H2} it was shown that $\fcb(M^0)
= \{Q^0_{x^0} \st x^0 \in K^0\}$.  

To see that any $Q_{x^0}$ is an IP, pick some point $x'
\in K'$, and consider the curve $\sigma(t) = (t,x^0,x')$;
we need to show $Q_{x^0} = I^-[\sigma]$.  Surely anything
to the past of $\sigma$ lies in
$Q_{x^0}$ (a timelike curve in
$M$ from $(s,y^0,y')$ to $(t,x^0,x')$ projects to a
timelike curve in $M^0$ from $(s,y^0)$ to $(t,x^0)$). 
For the converse, showing that $Q_{x^0}$ lies in the past
of $\sigma$, consider any $(s,y^0,y') \in Q_{x^0}$.  We
know from Proposition 5.2 $Q^0_{x_0} = I^-[\sigma^0]$ in
$M^0$, where $\sigma^0$ is the curve $\sigma^0(t) =
(t,x^0)$.  This tells us $(s,y^0)$ is in the past of
some $(t,x^0)$ in $M^0$; hence, $(s,y^0,y') \ll
(t,x^0,y')$ in $M$.  The trick is to show $(t,x^0,y')
\ll (t',x^0,x')$ for some $t' < b$; this is
where we use $\int_{b^-}^b f_j^{-1/2} = \infty$ for each
$j > k$. 

Consider any $j > k$, and, as before, let $M_j = (a,b)
\times_{f_j} K_j$.  Since $\int_{b^-}^b f_j^{-1/2} =
\infty$, $M_j$ is conformal to the product static
spacetime $\bar M_j = (a',\infty)\times K_j$.  For any
$p,q \in K_j$, for any $\tau > a'$, $(\tau',p) \gg
(\tau,q)$ so long as $\tau' > \tau + d_j(p,q)$, where
$d_j$ is the distance function from the Riemannian metric
on $K_j$.  Translating back to $M_j$ (which has the same
causal structure), we see that for any $p,q \in K_j$, for
any $s \in (a,b)$, there is some $s' \in (a,b)$ with
$(s',p) \gg (s,q)$.  We then apply this to $s = t$, $p =
x'_{k+1}$, and $q = y'_{k+1}$ to obtain a $t'_1$ with
$(t,y'_{k+1}) \ll (t'_1,x'_{k+1})$ in $M_{k+1}$.  This
gives us, in $M$, $(t, x^0, y'_{k+1}, y'_{k+2}, \dots,
y'_m) \ll (t'_1, x^0, x'_{k+1}, y'_{k+2}, \dots, y'_m)$;
in other words, we have shown we can travel futurewards
from $y'$ towards $x'$, changing only in the first factor
(in $K_{k+1}$).  Repeating successively in the other
factors, we end up with some $t' = t_{m-k}'$ so that
$(t,x^0,y') \ll (t',x^0,x')$, as desired.

We need to know that these $Q_{x^0}$ are the only
elements of $\fcb(M)$.  Let $\sigma$ be any
future-endless timelike curve in $M$; we can assume
$\sigma(t) = (t, c^0(t), c'(t))$ for curves $c^0$ and $c'$
in, respectively, $K^0$ and $K'$.  We need to find $x^0
\in K^0$ such that $I^-[\sigma] = Q_{x^0}$.  

Consider the projection $\sigma^0$ of $\sigma$ to
$M^0$.  From Proposition 5.2 of \cite{H2}, $I^-[\sigma^0]
= Q^0_{x^0}$, where $x^0$ is a point in $K^0$ that $c^0$
approaches.  We need to see that $I^-[\sigma] = Q_{x^0}$,
\ie, $Q^0_{x^0} \times K'$.  For any point $(t,y^0,y') \ll
\sigma(s)$, we also have, in $M^0$, $(t,y^0) \ll
\sigma^0(s)$, \ie, $(t,y^0) \in Q^0_{x^0}$, so
$(t,y^0,y') \in Q^0_{x^0} \times K'$.  Conversely,
consider any $(t,y^0) \ll \sigma^0(s)$ (any $s$) and any
$y' \in K'$.  It will suffice to show there is some $s'$
with $(\sigma^0(s), y') \ll (\sigma^0(s'),c'(s'))$; this
is where the compactness of $K_j$ for $j > k$ comes into
play. 

We need to know how points $(s,z^0,z')$ and $(r,w^0,w')$
in $M$ are timelike-related; this requires a curve $\rho
: [s,r]\to M$ of the form $\rho(t) = (t, e^0(t),
e'(t))$ for curves $e^0 :[s,r] \to K^0$ and
$e': [s,r] \to K'$ satisfying 
$$1 > \sum_{i=1}^k f_i(t)|\dot e^0_i(t)|^2_i\; +
\sum_{j=k+1}^m f_j(t)|\dot e'_j(t)|^2_j\;\tag{$*$}$$ 
with $e^0$ going from $z^0$ to $w^0$ and $e'$ going
from $z'$ to $w'$.  Let us first ignore the $K^0$ factor. 
The crucial fact in the $K'$ factor is this:  For given
$s$, for any $\delta > 0$, for any $z', w' \in K'$, there
is some $r \in (s,b)$ and a curve $e':[s,r] \to K'$ from
$z'$ to $w'$ so that for each $j > k$, $f_j(t)|\dot
e'_j(t)|^2_j < \delta^2$ for all $t \in [s,r]$.  For if we
insist that $|\dot e'_j(t)|_j < \delta f_j(t)^{-1/2}$,
then the length $L_j$ of $e'_j$ is bounded by
$\delta\int_s^r f_j(t)^{-1/2}\,dt$; but since $\int_s^b
f_j^{-1/2} = \infty$, this bound can be as large as we
like, by suitable adjustment of $r$ (still keeping $r <
b$).  Since each $K_j$ is compact, there is some maximum
length that will suffice for all $L_j$, $k+1 \le j \le
m$, no matter what the points $z'_j$ and $w'_j$ are (just
choose the maximum diameter among the $K_j$).  Thus, some
$r<b$ will accommodate all the curves $e'_j$ on the
interval $[s,r]$, going from wherever to wherever. 
In effect, we can entirely ignore the second term in
inequality ($*$), as with proper choice of $r$, a curve
$e' :[s,r] \to K'$ will exist from $z'$ to $w'$ that
makes that term $<(m-k)\delta^2$.  Thus, given $s \in
(a,b)$, $z^0, w^0 \in K^0$, and $z',w' \in K'$, there is
some $r < b$ with $(s,z^0,z') \ll (r,w^0,w')$ in $M$ if
and only if there is some $r < b$ with $(s,z^0) \ll
(r,w^0)$ in $M^0$.  It follows that $I^+((s,z^0,z'))$
includes all points of the form $(r,w^0,w')$ such that
$(r,w^0) \gg (s,z^0)$ and $r$ is sufficiently large (but
still $< b$).

So let us now examine the question of whether there is
some $s'$ with $(\sigma^0(s), y') \ll (\sigma^0(s'),
c'(s'))$.  We know now that $I^+(\sigma^0(s), y') =
I^+(s,c^0(s),y')$ includes all points of the form
$(r,w^0,w')$ with $(r,w^0) \gg (s,c^0(s))$ and $r$
sufficiently large.  This includes all points of the form
$(\sigma^0(s'),w') = (s',c^0(s'),w')$ for $s'$
sufficiently large, since $\sigma^0(s') \gg \sigma^0(s)$
for $s' > s$.  And in particular, this includes
$(\sigma^0(s'),c'(s'))$ for $s'$ sufficiently large. 
Therefore, $I^-[\sigma] = Q_{x^0}$.

Now that we know $\fcb(M) = \{Q_{x^0} \st x^0 \in K^0\}$,
we need to know that these elements are distinct for
different $x_0$.  But this is easy, since $Q_{x^0} =
Q^0_{x^0} \times K'$ and from Proposition 5.2 in
\cite{H2}, $Q^0_{x^0} \neq Q^0_{y^0}$ for $x^0 \neq
y^0$.  Thus, we have fully identified $\fcb(M)$ with
$K^0$ and, hence, $\M$ as the chronological set $\bar M$,
with $\fcb(M)$ sitting there as $\{b\} \times K^0 \times
\{*\}$.  We still must verify that the topology on $\M$
is that claimed.

First note we can rely on Proposition 5.2 to show us that
$\fcb(M)$ is spacelike:  If $Q^0_{x^0} \times K'
\subset Q^0_{y^0} \times K'$, then $Q^0_{x^0} \subset
Q^0_{y^0}$, and Proposition 5.2 then tells us $x^0 =
y^0$.  As it is evident that we cannot have $Q_{x^0}
\subset I^-((t,y^0,y'))$, that shows all we need.

To show the topology of $\M$ is that of $\bar M$ first
consider a sequence $\sigma$ in $\fcb(M)$, \ie, $\sigma(n)
= Q_{x^0_n}$ for some sequence $\{x^0_n\}$ in $K^0$.  We
want to show $Q_{x^0} \in L(\sigma)$ if and only if $[x^0]
= \lim [x^0_n]$.  Since the boundary is spacelike, we have
$Q_{x^0} \in L(\sigma)$ if and only if for all $(t,y^0,y')
\in Q_{x^0}$, eventually $(t,y^0,y') \in Q_{x^0_n}$. 
Since $Q_{z^0} = Q^0_{z^0} \times K'$, this is equivalent
to having for all $(t,y^0) \in Q^0_{x^0}$, eventually
$(t,y^0) \in Q^0_{x^0_n}$.  By Proposition 5.2, we
know that is equivalent to $x^0 = \lim \{x^0_n\}$, and
that is equivalent to convergence in $\bar M$, \ie,
$[x^0] = \lim\{[x^0_n]\}$.  (Note that we don't have to
worry about an element of $M$ being in $L(\sigma)$,
since $\fcb(M)$ is closed, $M$ being a spacetime.)

The last thing to consider is a sequence $\sigma$ in $M$,
$\sigma(n) = (t_n,x^0_n,x'_n)$; we want to show 
$Q_{x^0} \in L(\sigma)$ is equivalent to $[x^0] =
\lim\{\sigma(n)\}$, \ie, $\lim\{t_n\} = b$ and
$\lim\{x^0_n\} = x^0$.  We have $Q_{x^0} \in L(\sigma)$
if and only if for all $(t,y^0,y') \in Q_{x^0}$,
eventually $(t,y^0,y') \ll (t_n,x^0_n,x'_n)$, and that is
equivalent to having for all $(t,y^0) \in Q^0_{x^0}$ and
all $y' \in K'$, eventually $(t,y^0,y') \ll
(t_n,x^0_n,x'_n)$.  In particular, this implies that for
all $t<b$ and any $y' \in K'$, eventually $(t,x^0,y') \ll
(t_n,x^0_n,x'_n)$, which implies eventually $(t,x^0) \ll
(t_n,x^0_n)$ in $M^0$; and by Proposition 5.2, that
implies $\lim\{t_n\} = b$ and $\lim\{x^0_n\} = x^0$. 
Conversely, suppose $\lim\{t_n\} = b$ and $\lim\{x^0_n\}
= x^0$.  By Proposition 5.2, we know $Q^0_{x^0} \in
L(\{(t_n,x^0_n)\})$, so for all $(t,y^0) \in Q^0_{x^0}$,
eventually $(t,y^0) \ll (t_n,x^0_n)$ in $M^0$.  But by
the analysis above for timelike relations in $M$, we know
that for $t$ fixed, if $t_n$ is sufficiently close to
$b$, then $(t,y^0) \ll (t_n,x^0_n)$ is sufficient to
imply for any $z',w' \in K'$, $(t,y^0,z') \ll
(t_n,x^0_n,w')$; and, in particular, that for any $y' \in
K'$, $(t,y^0,y') \ll (t_n,x^0_n,x'_n)$.  Since $\{t_n\}$
approaches $b$, we have this; therefore, $Q_{x^0} \in
L(\{(t_n,x^0_n,x'_n)\})$.
\qed \enddemo

As examples, consider the quasi-Kasner spacetimes,
$K = (0,\infty) \times_{f_1}  K_1 \times_{f_2} K_2
\times_{f_3} K_3$, where each $K_i$ is
one-dimensional---either $\Bbb R$ or a circle---and
$f_i(t) = t^{2p_i}$.  We have $\int_1^\infty f_i^{-1/2} <
\infty$ if and only if $p_i > 1$.  Thus, if all $p_i \le
1$ (as in the classical Kasner spacetimes, which also
require the sums of $\{p_i\}$ and of $\{p_i^2\}$ each to
be 1), then $\fcb(K)$ has null relations unless all three
of the factors are circles, in which case the boundary is
a single point.  (This corrects a misstatement in section
5.2 of \cite{H2}.)  If $p_1 > 1$ while $p_2
\le 1$ and $p_3 \le 1$, then $\fcb(K)$ has null relations
unless $K_2$ and $K_3$ are circles; in that case, the
boundary is $K_1$.  And so on for other combinations; we
see that the boundary can be spacelike and any of a
point, a line, a circle, a plane, a cylinder, a torus,
$\R3$, a plane cross a circle, a line cross a torus, or a
3-torus, depending on the exponents and the spacelike
factors. 

With the structure of the boundary of multi-warped
spacetimes in hand, we can see how they fit into Theorem
3.4.  The interesting fact is that no extra assumption
about $\V/G$ need be made:

\proclaim{Proposition 3.6} Let $M$ be a multi-warped
spacetime covered by a spacetime $V$ supporting a free,
properly discontinuous, chronological group action from a
group $G$ so that $M = V/G$.  Suppose  $V$ and $M$ have
only spacelike future boundaries.    Then $\M$ is
homeomorphic  $\V/G$, and $\fcb(M)$ is homeomorphic to
$\fcb(V)/G$, \ie, $\widehat{V/G} \cong \V/G$ and
$\fcb(V/G) \cong \fcb(V)/G$. \endproclaim

\demo{Proof} We have $M = (a,b) \times_{f_1} K_1 \cdots
\times_{f_m} K_m$.  Arrange the factors so that
for some $k$,  $\int_{b^-}^b f_i^{-1/2} <
\infty$ for $i \le k$ and $\int_{b^-}^b f_i^{-1/2} =
\infty$ for $i > k$; then, by Proposition 3.5, each
$K_i$ is complete, and for $i>k$, $K_i$ is
compact.  

It is easy to work out what $V$ is:  Since $V$
covers $M$, $V$ is a quotient of the universal cover for
$M$, $\widetilde M$; in fact, $G$ must be a  normal
subgroup of the fundamental group of $M$, $H = \pi_1(M)$,
and $V = \widetilde M/(H/G)$.  Nor is there any difficulty
in determining what $\widetilde M$ is:  It must be $(a,b)
\times_{f_1} \widetilde K_1 \cdots \times_{f_m} \widetilde K_m$
(with $\widetilde K_i$ denoting the universal cover of
$K_i$), as this is plainly simply connected and also a
cover of $M$.  Evidently, $H = H_1 \times \cdots
\times H_m$, where $H_i = \pi_1(K_i)$ and $H$ acts on
$\widetilde M$ via $(h_1, \dots, h_m) \cdot (t,  x_1,
\dots, x_m) = (t, h_1 \cdot x_1, \dots, h_m
\cdot x_m)$.  Since $G$ is a subgroup of the product
group $H$, $G$ must also break down with factors $G =
G_1 \times \cdots \times G_m$, each $G_i$ a subgroup of
$H_i$; furthermore, as $G$ is normal in $H$, each $G_i$
is normal in $H_i$, and $H/G \cong (H_1/G_1) \times
\cdots \times (H_m/G_m)$.  It follows that $V = (a,b)
\times_{f_1} \bar K_1 \cdots \times_{f_m} \bar K_m$, where
$\bar K_i = \widetilde K_i/(H_i/G_i)$.  Completeness and the
integrals of the warping functions are unaffected by all
these covers and quotients, but compactness may not be. 
For $V$ to have a spacelike future boundary we must have
$\bar K_i$ compact for all $i>k$.

By Proposition 3.5, $\fcb(V) \cong \bar K_1 \times \cdots
\times \bar K_k$.  The action of $G$ on $\fcb(V)$ derives
from the action of $H$ on $\widetilde M$, so $(g_1, \dots, 
g_m) \cdot (x_1, \dots, x_k) = (g_1\cdot x_1, \dots,
g_k\cdot x_k)$ (with $g_{k+1}$ through $g_m$
irrelevant).   This action is not free (if $k < m$), but
we can analyze it in terms of $G^0 = G_1 \times \cdots
\times G_k$, which does act freely on $\fcb(V)$, since
$G$ acts freely on $V$ (as that implies each $G_i$ acts
freely on $K_i$).  What is particularly important is
that $G^0$ acts properly discontinuously on $K^0$.

Now consider an element of
$\fcb(V)/G^0$:  It is an equivalence class
$\<Q_{x^0}\>$ for some $x^0 \in K^0 = K_1 \times \cdots
K_k$, under the equivalence relation of movement by
elements of $G$ (using the $Q$-notation of the proof of
Proposition 3.5).  The past of $\<Q_{x^0}\>$ in the
chronological set $\V/G$ is $\{\<p\> \st p \in
Q_{x^0} \}$, \ie, the $G$-orbits of all points in
$Q_{x^0}$.  Suppose $\<Q_{y^0}\>$ has the same past as
$\<Q_{x^0}\>$; then all $G$-orbits of points in $ Q_{x^0}$
are also $G$-orbits of points in $Q_{y^0}$, and {\it vice
versa\/}.  In particular, for all $t<b$, for any $x' \in
K' = K_{k+1} \times \cdots \times K_m$, there is some $g
\in G$ such that $g\cdot(t,x^0,x') \in Q_{y^0}$.  This
amounts to saying that for all $n$, there is some $g^0_n
\in G^0$ such that (for $b$ finite) $(b - \frac1{n},
g^0_n\cdot x^0) \in Q^0_{y^0}$ or (for $b = \infty$) $(n,
g^0_n\cdot x^0) \in Q^0_{y^0}$ (using the $Q^0$-notation
from the proof of Proposition 3.5).  The proof of
Proposition 5.2 shows that the $s$-slices---\ie,
intersections with $\{s\} \times K^0$---of each
$Q^0_{z^0}$ narrow down to $\{z^0\}$ as $s$ approaches
$b$ (that is how it was shown that $Q^0_{z^0} = Q^0_{w^0}$
implies $z^0 = w^0$).  Thus we have $\{g^0_n\cdot x^0\}$
approaches $y^0$.  Since $G^0$ acts properly
discontinuously on $K^0$, this can happen only if $x^0$
and $y^0$ are in the same $G^0$-orbit.  Therefore,
$I^-(\<Q_{x^0}\>) = I^-(\<Q_{y^0}\>)$ implies $\<x^0\> =
\<y^0\>$, which implies $\<Q_{x^0}\> = \<Q_{y^0}\>$: 
$\widehat{V/G}$ is past-distinguishing.  By Theorem 3.4,
$\widehat{V/G} = \V/G$ and $\fcb(V/G) = \fcb(V)/G$. \qed 
\enddemo

Proposition 3.6 shows that when a multi-warped
spacetime is covered, via group action, by another
spacetime, the covering spacetime is also multi-warped;
and then the behavior of the boundaries, if spacelike,
is very simple.  But what if we start with the covering
spacetime being multi-warped; does it follow that its
quotient by a group action is also multi-warped?  In
other words:  When examining $V \to M$, a principal
covering projection, we now know that $M$ being
multi-warped propagates up the projection to the same
being true for $V$; but what about the converse, with $V$
being assumed multi-warped?

If no assumptions about the spacelike nature of the
boundaries are made, the group action on $V$ can be more
complicated than that established in Proposition 3.6. 
Consider, for instance, $V = \mk1 \times_f \R1$, with
$f(t)$ positive but bounded above (so $V$ has a null
boundary).  We can act isometrically on $V$ by the
integers $\Bbb Z$ via $m \cdot (t,x) = (t+\frac12m,
x+m)$, so long as $f$ is periodic with period $\frac12$;
this will be a chronological action so long as
$f(t)$ is sufficiently close to 1 for all $t$.  Then
$V/\Bbb Z$ is a strongly causal spacetime but does not
have the structure of a multi-warped spacetime unless $f$
is very specially chosen:  For a multi-warped spacetime
will be of the form $\mk1 \times_\phi \Sph1_D$ for some
function $\phi$ with $\Sph1_D$ the circle of diameter
$D$.  As $\mk1 \times_\phi \Sph1_D$ contains a closed
spacelike geodesic of length $D$, so must $V/\Bbb Z$ if
it is multi-warped, which translates into a spacelike
geodesic in $V$ which is preserved by the $\Bbb Z$-action.
That means $D$ must be the spacelike separation between $p
= (0,0)$ and $q = (\frac12,1)$, and a preserved geodesic
must go between those points.  But while there is surely a
spacelike geodesic $\gamma$ from $p$ to $q$, only a
special choice of $f$ will allow for $\dot\gamma(D)$ to
be the image of $\dot\gamma(0)$ by the translation 
$(t,x) \mapsto (t+\frac12,x+1)$. 

Does imposing the restriction of a spacelike boundary
alter things for the simpler?  If the timelike factor for
$V$ is future-finite, \ie, if the $(a,b)$ factor has $b$
finite, then the answer is yes:  Any isometry of $V$ must
preserve the length of the longest timelike curve issuing
from a point.  From a point $(t,x)$, that length is
$b-t$, so the isometry must preserve the $\{t\}$-slice in
$V$.  Similarly, the length of the longest timelike curve
between any two points must be preserved, so if the
isometry takes $(t,x)$ to $(t,y)$, then it must take
$(s,x)$ to $(s,y)$, as $(s,y)$ is the only point in the
$\{s\}$-slice with longest curve-length $|t-s|$ from
$(t,y)$.  Thus, any group acting by isometries on $V$
must act in the manner described in Proposition 3.6, and
the quotient $M$ is again multi-warped.  Then
Proposition 3.6 applies.  

But what if $V$ has spacelike boundary with $b=\infty$? 
That is unclear.  No examples with irregular group action
have come to light to date, but a proof that the group
action must be as above seems elusive.

\subhead Example with $\hat\pi$ not continuous
\endsubhead

Here is an example demonstrating what can go wrong in the
absence of spacelike boundaries.

Consider the following:  Let $M$ be the open subset of
$\mk2$ formed by removing an infinite number of closed
timelike line segments:  Let $L^-$ and $L^+$ be
respectively the null lines $\{t=x\}$ and $\{t=x+3\}$. 
Let $\sigma^-$ and $\sigma^+$ be timelike curves
respectively asymptotic to $L^-$ and $L^+$, $\sigma^-(s)
= (\cosh s,\sinh s)$ and $\sigma^+(s) = (\cosh s,(\sinh
s)+3)$.  For each positive integer $n$, let $S_n$ be the
closed segment from $\sigma^-(n)$ to $\sigma^+(n)$.  Let
$M = \mk2 - \bigcup\{S_n \st n \ge 1\}$.  Let $V =
\widetilde M$, the universal cover of $M$, and let $G =
\pi_1(M)$, the fundamental group of $M$ acting on $V$ so
that $V/G = M$.  Let $\pi: V \to M$ be the projection.

The elements of $G$ are the homotopy classes of loops in
$M$ based at, say, $(0,0)$.  These can be realized as
ordered $n$-tuples of integers, $s = (s_1,
\dots, s_n)$, representing the sequence of slits around
which the loop travels, with
positive entries representing counterclockwise travel
around the indicated slit ($i$th slit is $S_k$ if $s_i
= k > 0$) and negative numbers clockwise
travel about the slit ($S_k$ for $s_i = -k < 0$); the
empty sequence is the identity element.  We have the
identifications $s
\equiv s'$, in case
$s_i = -s_{i+1}$ and $s'$ is $s$ with $s_i$ and
$s_{i+1}$ deleted.  The group operation is
 concatenation of sequences.

$V$ can be realized as an infinite collection of
identical sheets, parametrized by $G$, with gluings
among them according to $G$.  Specifically, let $N$ be
the manifold obtained from $M$ by deleting horizontal
slits running from each $S_n$ to $S_{n+1}$:  Let $H_n =
\{(x,(\sinh n)+2) \st \cosh n<x<\cosh(n+1)\}$; then $N = M
- \bigcup\{H_n \st n \ge 1\}$.  Note that $N$ is
homeomorphic to the plane, since it is
$\mk2$ with a connected closed tree-like set deleted.  We
realize $V$ as $N \times G$ with the sheet $N \times
\{s\}$ glued to the sheet $N \times \{s \cdot (k)\}$ 
across $H_k$ (points on upper edge of $H_k$ in
$N\times\{s\}$ connected to points on lower edge of $H_k$
in $N\times\{s\cdot(k)\}$). 

Let $P$ and $Q$ be the IPs in $M$ generated
respectively by the future chains $c^+
= \{\sigma^+(n+1/2)\}$ and $c^- = \{\sigma^-(n+1/2)\}$. 
Actually, $P = I^-[L^+]$, but
$Q$ does not have so neat a formulation, as portions of
the slits $\{S_n\}$ keep it from extending to $L^-$; but
the boundary of $Q$ is asymptotic to $L^-$.  (Note that
$P$ properly contains $Q$; this is the crucial
non-spacelike aspect of the boundary in $M$.)  Let
$\widetilde c^+$ and $\widetilde c^-$ be pre-images under $\pi$,
respectively, of $c^+$ and $c^-$ in $N\times\{e\}$ ($e
= (\,)$, the identity element in $G$), and let $\tP =
I^-[\widetilde c^+]$ and $\tQ = I^-[\widetilde c^-]$, IPs in $V$
mapping via $\pi$ onto $P$ and $Q$ respectively; $\tQ$
maps homeomorphically to $Q$, but $\tP$ is more
complicated, as it extends across each $H_k$ into
$N\times\{(k)\}$, where it occupies a portion of the IP
in $N\times\{(k)\}$ mapping homeomorphically onto
$Q$---that portion with $t < -x+\cosh(k+1)$.  

Just as $\tP$ extends across $H_k$ in $N\times\{e\}$ to a
portion of the pre-image of $Q$ lying in $N\times\{(k)\}$
(\ie, $(k)\cdot\tQ$), so $\tP_k = (-k)\cdot
\tP$ extends across $H_k$ in $N\times\{(-k)\}$ to a
portion of $\tQ$---that portion with $t < -x +
\cosh(k+1)$, points to the past of the null line through
$\widetilde c(k+1)$ rising to the left.  Accordingly, every
point in $\tQ$ is  contained in $\tP_k$ for $k$
sufficiently large.  Furthermore, for every IP $I$
properly containing $\tQ$, there is some point in $I$ that
fails to lie in $\tP_k$ for $k$ sufficiently large
(because any IP properly containing $\tQ$ must contain,
for some $\delta > 0$, $\{(x,t) \in N\times\{e\} \st
t<x+\delta\}$).  This says precisely that $\tQ$ is a
limit, in the \htop of $\V$, of the sequence $\{\tP_k \st
k\ge 1\}$.

We have $\hat\pi(\tP_k)= P$ for each $k$ and
$\hat\pi(\tQ) = Q$.  Thus, if $\hat\pi$ were continuous,
the constant sequence $\{P\}$ would have to have $Q$ as a
limit in the \htop of $\M$.  But as $M$ is
past-distinguishing (being strongly causal), so is $\M$
(by Theorem 5 of \cite{H1}), so Proposition 2.1 of
\cite{H2} tells us that points in
$\M$ are closed; thus, the only limit of the constant
sequence $\{P\}$ is $P$.  (More concretely: $P$ is an IP
properly containing $Q$ such that every element of $P$ is
eventually contained in the elements of the constant
sequence $\{P\}$; thus $Q$ is not a limit of that
sequence.)  Therefore, $\hat\pi:\V\to\M$ is not
continuous.

(This is also an example of $\Pi : \V \to (\V/G,
\text{\htop})$ being not continuous:  Since each of the
$\tP_k$ are $G$-related to $\tP$, we have for all $k$,
$\Pi(\tP_k) = \<\tP_k\> = \<\tP\>$, so $\{\Pi(\tP_k)\}$
is the constant sequence $\{\<\tP\>\}$.  But just as $P$
is an IP in $M$ properly containing $Q$, so is $\tP$ an
IP in $V$ properly containing $\tQ$, and so is $\<\tP\>$
an IP in $V/G$ properly containing $\<\tQ\>$.  Therefore,
$\Pi(\tQ) = \<\tQ\> \not\in L(\{\Pi(\tP_k)\})$, even
though $\tQ \in L(\{\tP_k\})$.)

\end